\documentclass[showpacs,amsmath,amssymb,aps,longbibliography,twocolumn,pra]{revtex4-2}
\usepackage{graphicx}
\usepackage{soul}
\usepackage[colorlinks=true,citecolor=blue,linkcolor=magenta]{hyperref}
\usepackage[usenames]{color}
\usepackage[english]{babel}
\usepackage{enumerate}
\usepackage{pifont}
\usepackage{amsfonts}
\usepackage{siunitx}

\usepackage{braket}
\usepackage{bbm}

\usepackage[dvipsnames]{xcolor}
\usepackage{cancel,comment} 
\usepackage[normalem]{ulem} 

\newcommand{\figref}[1]{Fig.\,\ref{#1}}
\renewcommand{\eqref}[1]{Eq.\,\ref{#1}}

\newcommand {\I} {\mathrm{i}}
\newcommand {\E} {\mathrm{e}}

\usepackage{xr}

\begin{document}

\title{Real-space detection and manipulation of topological edge modes with ultracold atoms}

\author{Christoph Braun$^{1,2,3}$}
\thanks{These two authors contributed equally}
\author{Rapha\"el Saint-Jalm$^{1,2,3}$}
\thanks{These two authors contributed equally}
\author{Alexander Hesse$^{1,2,3}$} 
\author{Johannes Arceri$^{1,2,3}$}
\author{Immanuel Bloch$^{1,2,3}$}
\author{Monika Aidelsburger$^{1,2}$}
\email{monika.aidelsburger@physik.uni-muenchen.de}
\affiliation{$^{1}$\,Fakult\"at f\"ur Physik, Ludwig-Maximilians-Universit\"at M\"unchen, Schellingstra{\ss}e 4, 80799 M\"unchen, Germany}
\affiliation{$^{2}$\,Munich Center for Quantum Science and Technology (MCQST), Schellingstra{\ss}e 4, 80799 M\"unchen, Germany}
\affiliation{$^{3}$\,Max-Planck-Institut f\"ur Quantenoptik, Hans-Kopfermann-Stra{\ss}e 1, 85748 Garching, Germany}


\maketitle

\textbf{
Conventional topological insulators exhibit exotic gapless edge 
or surface states, as a result of non-trivial bulk topological properties.
In periodically-driven systems the bulk-boundary correspondence is fundamentally modified 
and knowledge about conventional bulk topological invariants is insufficient.
While ultracold atoms provide excellent settings for clean realizations of Floquet protocols, 
the observation of real-space edge modes has so far remained elusive.
Here we demonstrate an experimental protocol for realizing chiral edge modes in optical lattices, 
by creating a topological interface using a potential step that is generated with a programmable optical potential. 
We show how to efficiently prepare particles in these edge modes 
in three distinct Floquet topological regimes that are realized in a periodically-driven honeycomb lattice. 
Controlling the height and sharpness of the potential step, 
we study how edge modes emerge at the interface and how the group velocity 
of the particles is modified as the sharpness of the potential step is varied. 
}

\def\thefootnote{*}\footnotetext{These authors contributed equally to this work}

\begin{figure*}[!htb]
\includegraphics{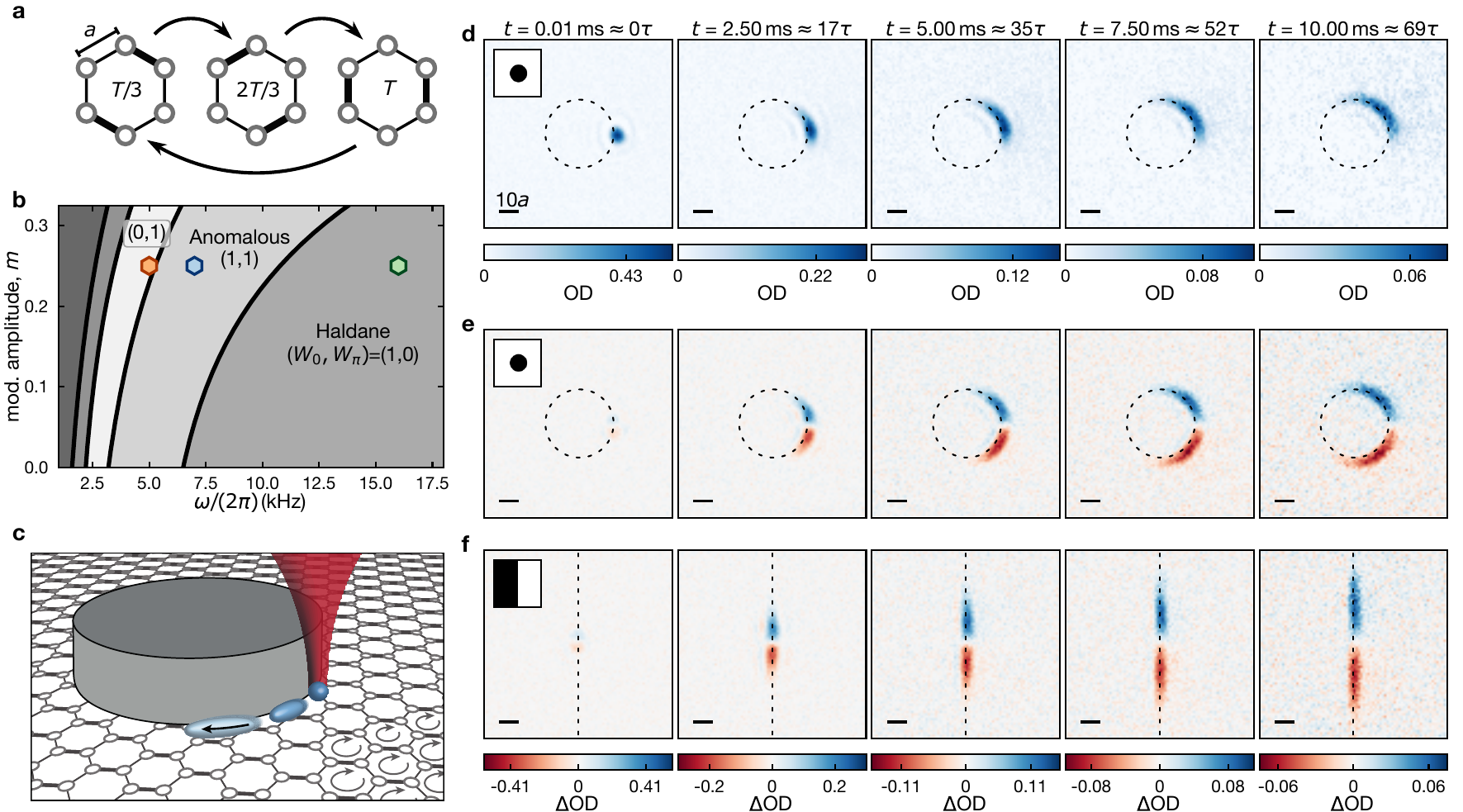}
\vspace{-0.cm} \caption{\textbf{Preparation and observation of anomalous Floquet topological edge states.}
  \textbf{a}, Sketch of the modulation scheme, where the hopping rate between neighboring sites is varied periodically in a chiral manner (arrows) with modulation period $T=2\pi/\omega$. The thick (thin) lines in the three snapshots indicate larger (smaller) tunnel coupling. The lattice spacing is $a=287\,\si{\nano\meter}$.
  \textbf{b}, Phase diagram of the amplitude-modulated honeycomb lattice. The different topological regimes are characterized by the tuple of winding numbers $(W_0,W_\pi)$ of the two quasienergy gaps. The hexagons mark the modulation parameters used in this work [$m=0.25$, green: $\omega/(2\pi)=16\, \si{\kilo\hertz}$, blue: $\omega/(2\pi)=7\, \si{\kilo\hertz}$ and orange: $\omega/(2\pi)=5\,\si{\kilo\hertz}$].
\textbf{c}, Illustration of the optical potential and the generation of the initial state. A potential step (gray cylinder) is applied on a selected region of the system to block the motion of particles in the modulated lattice (arrows), generating a sharp edge. The initial state is prepared by trapping a cloud of atoms (BEC, indicated in dark blue) in an optical tweezer (red) near the edge. After releasing the cloud, the atoms exhibit a chiral motion along the edge, illustrated by the different blue shadings.
\textbf{d}, Time-evolution of the atoms after releasing the atoms from the tweezer into the modulated lattice ($\kappa = +1$), close to a disk-shaped potential with radius $5.8\,\si{\micro\meter}$ ($\approx 20 a$) represented by the dashed line; OD denotes optical density. The inset illustrates the shape of the repulsive potential, which is shown in black. 
\textbf{e}, Difference image $\Delta\text{OD} = \text{OD}_{\kappa=+1} - \text{OD}_{\kappa=-1}$ between the time evolution shown in \textbf{d} with $\kappa=+1$ and the evolution under the opposite chirality ($\kappa=-1$) of the lattice modulation.
\textbf{f}, Same as \textbf{e} for a straight edge (dashed line). The corresponding potential is shown in the inset.
The evolution times of \textbf{d}-\textbf{f} are indicated at the top of each column, the tunnel coupling of the static lattice is $J_0 = h\times1.1(1)\,\si{\kilo\hertz}$ (tunnelling time $\tau = \hbar/J_0 = 145\,\si{\micro\second}$) and the modulation parameters are $\omega=2\pi \times 7\,\si{\kilo\hertz}$ and $m=0.25$ (blue marker in \textbf{b}).
All absorption images are averages of $100-300$ individual experimental realizations. The scale bar represents a length of $10a$. The position of the dashed lines is calculated from the programmed pattern on the DMD.
}
\label{Fig_1}
\end{figure*}


Topological and geometrical properties of wave functions are 
at the heart of many intriguing electronic properties of materials~\cite{xiao_berry_2010}. 
The most prominent examples are topological insulators, superconductors, and superfluids~\cite{hasanColloquiumTopologicalInsulators2010,bernevigTopologicalInsulatorsTopological2013,qi_topological_2011}.
A remarkable concept regarding their properties is the bulk-boundary correspondence, which links the topological properties of the bulk to the existence of exotic gapless states at the edge or surface of the system.

The manifestation of the bulk-boundary correspondence is most easily understood from the integer quantum Hall effect~\cite{klitzing_new_1980,von_klitzing_quantized_1986}, whose characteristic transport properties exhibit two main features:
a precise quantization of the transverse Hall conductivity determined by the Chern number~\cite{thoulessQuantizedHallConductance1982,niu_quantised_1984} -- 
a topological invariant characterizing the bulk energy bands -- and the existence of robust chiral edge modes at the boundary.
According to the bulk-boundary correspondence~\cite{halperin_quantized_1982,rammal_quantized_1983,macdonald_edge_1984,hatsugai_chern_1993,hatsugai_edge_1993,qi_general_2006}, knowledge of the bulk Chern numbers is sufficient to predict the number and chirality of the edge modes.

Synthetic quantum systems have emerged as promising experimental routes for exploring the rich properties of topological systems~\cite{aidelsburger_artificial_2018,cooperTopologicalBandsUltracold2019,ozawaTopologicalPhotonics2019}. 
While many of the concepts described above are well understood both in theory and experiment, engineered quantum systems offer new possibilities for generating exotic topological states that a priori have no analogue in solid-state systems.
This was recently demonstrated in periodically-driven systems with photonic waveguides~\cite{mukherjeeExperimentalObservationAnomalous2017a, maczewskyObservationPhotonicAnomalous2017a}, resonator arrays~\cite{pengExperimentalDemonstrationAnomalous2016,gao_probing_2016,afzalRealizationAnomalousFloquet2020} and with cold atoms~\cite{winterspergerRealizationAnomalousFloquet2020}.
There, due to the time-periodic nature of the Hamiltonian, a generalized form of the bulk-boundary correspondence predicts the existence of chiral edge modes even in situations where the Chern number of the bulk band vanishes~\cite{kitagawaTopologicalCharacterizationPeriodically2010,rudnerAnomalousEdgeStates2013}.

The generalized bulk-boundary correspondence strongly motivates 
the development of experimental techniques that enable both the study of bulk 
and edge topological properties in synthetic quantum systems to reveal their connection.
In cold atoms  a large variety of techniques has been developed to study bulk geometrical properties using interferometric or state-tomography techniques~\cite{atala_direct_2013,duca_aharonov-bohm_2015,flaschner_experimental_2016,li_bloch_2016},  transport measurements~\cite{jotzu_experimental_2014,aidelsburger_measuring_2015,zhou_observation_2022},  as well as methods based on spectroscopy or quench dynamics~\cite{tarnowski_measuring_2019,asteria_measuring_2019}. 

The observation of edge modes in photonic devices is facilitated by a natural sharp boundary~\cite{hafezi_imaging_2013,rechtsman_photonic_2013,ozawaTopologicalPhotonics2019}.
With cold atoms, their observation was enabled by the concept of synthetic dimensions~\cite{celi_synthetic_2014,ozawa_topological_2019}.
There, one real-space dimension is replaced by internal degrees of freedom, 
and the finite number of coupled internal levels naturally creates a well-defined boundary~\cite{manciniObservationChiralEdge2015,stuhlVisualizingEdgeStates2015,chalopinProbingChiralEdge2020}.
In real-space 2D systems, however, the edges of the system are typically smooth, hindering the observation of edge modes.
In order to overcome this limitation, several strategies have been proposed~\cite{goldmanDetectingChiralEdge2012, goldmanDirectImagingTopological2013,goldmanCreatingTopologicalInterfaces2016}, but their observation with cold atoms has so far remained elusive.

In this work, we generate a topological interface in a 2D real-space Floquet system~\cite{winterspergerRealizationAnomalousFloquet2020} 
by creating a potential step.
We demonstrate the presence of topological edge modes by preparing a localized Bose-Einstein condensate (BEC), 
which after release into the lattice near the edge exhibits a chiral motion characteristic of topological edge modes.
We further study the preparation of these edge modes in three distinct topological regimes 
as a function of the initial-state parameters, and show how the edge modes emerge as the parameters of the interface are modified.


The experimental setup consists of a BEC of $^{39}$K atoms. The optical honeycomb lattice is formed by three propagating, $s$-polarized, blue-detuned laser beams (wave vector $k_\mathrm{L} = 2\pi/\lambda_L$, with $\lambda_\mathrm{L} = 745\,\si{\nano\meter}$), which interfere in the $xy$-plane at relative angles of $120\si{\degree}$~\cite{winterspergerRealizationAnomalousFloquet2020}.
An additional harmonic confinement is provided by an optical dipole trap at $1064\,\si{\nano\meter}$ generated by two laser beams that cross at $90\si{\degree}$ in the $xy$ plane, resulting in a radial trapping frequency of $\omega_r = 2\pi \times 17(1) \,\si{\hertz}$, and a vertical trap frequency of $\omega_z = 2\pi \times 330(30)\,\si{\hertz}$.

The periodic modulation is realized by modulating the
intensity $I_i$ ($i=\{1,2,3\}$) of the individual lattice laser beams according to $I_i(t)=I_0(1- m + m\, \text{cos}(\omega t + \phi_i))$, where $\omega$ denotes the modulation frequency, $m$ the modulation amplitude, $\phi_i = \kappa \times \frac{2\pi}{3} i$ the phase for the $i$th beam and $\kappa = \pm 1$ the chirality.
This results in a chiral modulation of the tunnel couplings in the honeycomb lattice, as illustrated in \figref{Fig_1}a. The periodic modulation breaks time-reversal symmetry and results in a rich phase diagram with several topological regimes~\cite{winterspergerRealizationAnomalousFloquet2020}, as shown in \figref{Fig_1}b. 
The modulated lattice realizes an effective two-band model similar to Ref.~\cite{kitagawaTopologicalCharacterizationPeriodically2010}.
The corresponding bulk topological properties have been studied experimentally in Ref.~\cite{winterspergerRealizationAnomalousFloquet2020}
using local Hall deflection measurements with large BECs.

A full characterization of the topological properties in driven systems is given by the winding number in each gap, which determines both the number and chirality of the edge modes in the respective gap~\cite{kitagawaTopologicalCharacterizationPeriodically2010,rudnerAnomalousEdgeStates2013}.
The winding number is computed by taking the full time evolution of the system into account. Here, we denote the two winding numbers as $W_0$ and $W_\pi$, and the respective energy gaps as $0$- and $\pi$-gap; the latter is located at the edge of the Floquet Brillouin zone at quasienergy $\epsilon = \pm \hbar\omega/2$; $\hbar=h/(2\pi)$ is Planck's constant.
The conventional Chern numbers are linked to the winding numbers according to $\mathcal{C}^\pm = \pm (W_\pi - W_0)$, where $\mathcal{C^{-(+)}}$ denotes the Chern number of the lower (upper) energy band [$\mathcal{C^{-}}+\mathcal{C^{+}}=0$].
The three regimes studied in this work are displayed in \figref{Fig_1}b.
All of them exhibit topological edge modes, even in the anomalous regime where the Chern number of the bulk bands vanishes.

\begin{figure*}[!htb]\includegraphics{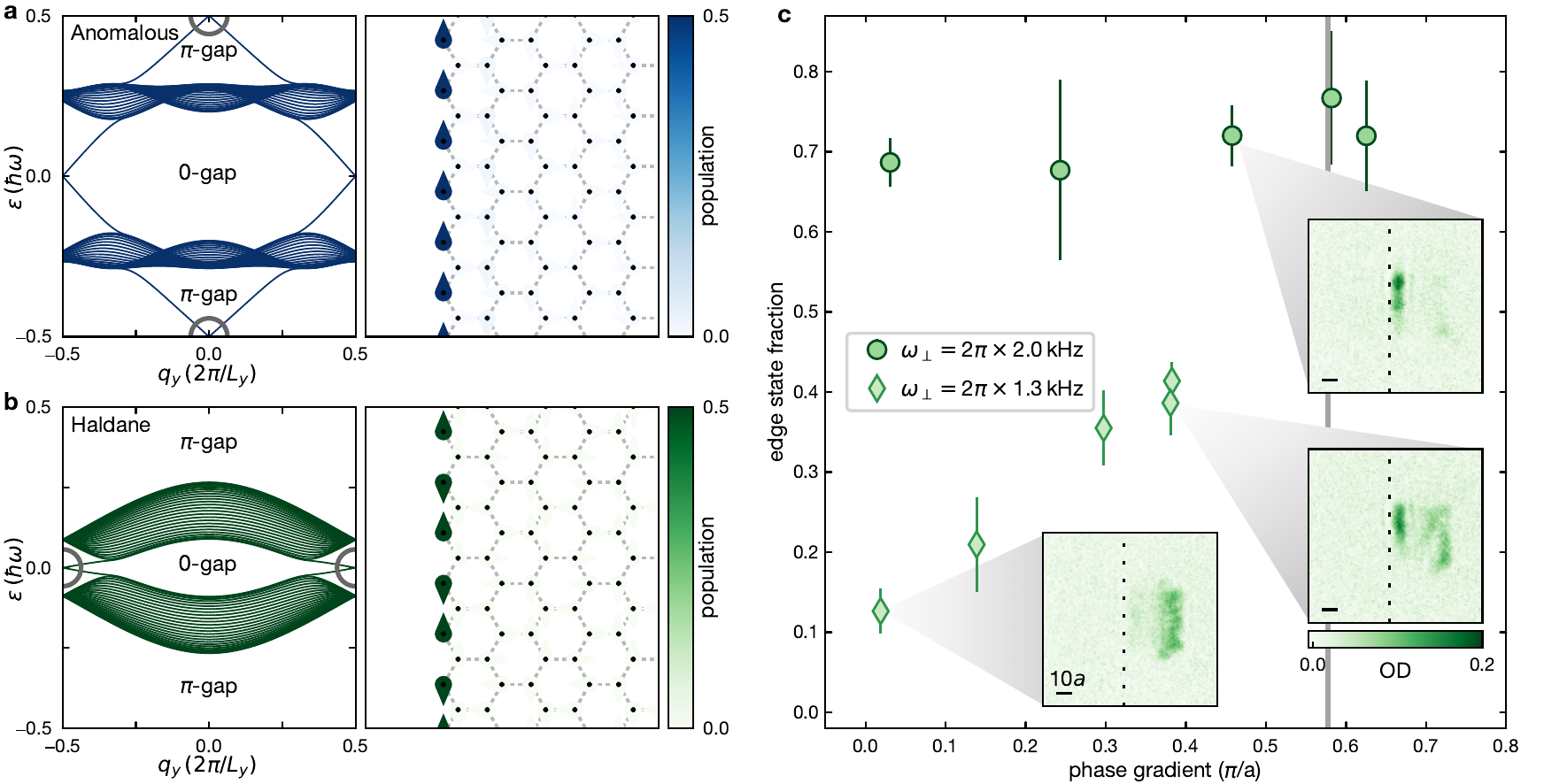}
\vspace{-0.cm}
\caption{\textbf{Preparation and observation of edge states in the Haldane regime}.
\textbf{a}, Band structure of the tight-binding model in the anomalous regime with $(W_0,W_\pi)=(1,1)$ and the corresponding wave function of the edge state in the $\pi$-gap at $q_y=0$, as highlighted by the gray circle; $\varepsilon$ denotes the quasienergy. The spectrum is plotted for the first Floquet Brillouin zone $-\hbar\omega/2<\varepsilon\leq\hbar\omega/2$.
\textbf{b}, Same as \textbf{a} for the Haldane regime with $(W_0,W_\pi)=(1,0)$.
The right panel shows the wave function of the edge state in the $0$-gap at $q_y = \pi/L_y = \pi/(\sqrt{3} a) \approx 0.58 \pi/a$, as highlighted by the gray circle.
For the panels on the right in \textbf{a} and \textbf{b}, the color shade of the arrows indicates the modulus square of the wave function
and their direction indicates the complex phase. 
\textbf{c}, Fraction of atoms in the edge mode after a time evolution of $1.5\,\si{\milli\second}$ as a function of the phase gradient applied to the initial wave packet (see Methods).
The data points are averaged from 4-5 realizations and the error bars represent the propagation of the background noise of the absorbtion image to the population fraction.
The gray vertical line indicates the edge of the Brillouin zone, where $q_y = \pi/L_y$.
The insets show averaged absorption images (88-90 realizations) of the atomic cloud after a time evolution of $3\,\si{\milli\second}$ in the Haldane regime. The color bar below is common to all insets. 
}
\label{Fig_2}
\end{figure*}

The experimental scheme used to probe topological edge states is schematically depicted in \figref{Fig_1}c.
The potential edge is realized by illuminating a selected area of the lattice with a programmable, binary, repulsive optical potential derived from an incoherent light source with a wavelength centered around $638\,\si{\nano\meter}$, and shaped with a digital micromirror device (DMD), as described in the Methods. 
The width of the potential edge has a lower bound of $\approx 0.7\,\si{\micro\meter}$~\cite{DefinedLength92}, which is set by the numerical aperture of the imaging system (see Methods). 
The actual width in the atomic plane is most likely slightly larger due to residual aberrations.
To prepare an initially localized wave packet about 200 atoms are adiabatically transferred from a BEC generated in the crossed dipole trap to a tightly focused optical tweezer [$\lambda_\mathrm{T} = 1064\,\si{\nano\meter}$; radial trap frequency $\omega_\perp = 2\pi\times 2.0(1) \,\si{\kilo\hertz}$]. The tweezer beam propagates perpendicular to the lattice plane and its position is controlled by an acousto-optic deflector (AOD).
Residual atoms outside the tweezer are expelled from the trap. The scattering length of the atoms is tuned to $a_\mathrm{s} = 6a_0$ utilizing the Feshbach resonance at $403\,\si{G}$, here $a_0$ denotes the Bohr radius. 

After the preparation of a tightly-confined BEC in the tweezer, the programmable repulsive potential is ramped up to its final height in $30\,\si{\milli\second}$ and subsequently the optical lattice is ramped up to a depth of $6E_\mathrm{r}$ in $30\,\si{\milli\second}$, where $E_\mathrm{r}=\frac{\hbar^2 k_\mathrm{L}^2}{2m_{\text{K}}}=  h \times 9.23\,\si{\kilo\hertz}$ is the recoil energy of the lattice and $m_\mathrm{K}$ denotes the mass of a $^{39}$K atom. 
Finally the amplitude of the modulation is increased during the first five modulation periods. 
Once the modulation has reached its final amplitude, the optical tweezer is abruptly switched off and the evolution of the atoms is observed. 
We separately confirmed that the evolution of the wave packet is coherent after releasing it into the static lattice without the edge potential, and that the initial localization of the wave packet is sufficient to occupy high quasi-momentum states in the Brillouin zone (see Methods).

We start by studying the evolution of the localized wave packet in the anomalous regime (blue marker in \figref{Fig_1}b), where edge modes are expected to exist, although the bands exhibit Chern numbers $\mathcal C^\pm = 0$. 
We generate a disk-shaped repulsive potential with a height of $V_0 = h \times 16.7(3) \,\si{\kilo\hertz} > \hbar \omega$. 
The position of the tweezer relative to the edge of the potential was optimized in order to maximize the fraction of atoms in the edge mode. 
After releasing the atoms from the tweezer we observe that the wave packet propagates along the potential boundary, following its curvature, as is characteristic for chiral edge states (\figref{Fig_1}d). 
Furthermore, even though the potential is repulsive, the atoms remain close to the edge, propagate over more than $10a$ and do not scatter into the bulk of the system, indicating a good overlap of the initial wave packet with the edge mode.
In addition, the wave packet disperses while propagating, due to the non-linear dispersion relation of the edge mode, as a result of the finite width of the edge~\cite{stanescuTopologicalStatesTwodimensional2010, buchholdEffectsSmoothBoundaries2012, goldmanDirectImagingTopological2013}. 
To highlight the chiral nature of the edge state, the modulation direction is inverted, thus changing the sign of the winding numbers and therefore the chirality of the edge state. 
Figure~\ref{Fig_1}e depicts the difference between the images taken for the two chiralities.
Apart from the change in the propagation direction, we observe no difference. 

The programmable repulsive potential offers almost arbitrary control over the shape and orientation of the edge.
In \figref{Fig_1}f we show experimental results for a repulsive potential that creates a straight line along a zigzag-type edge of the honeycomb lattice. 
Similar to \figref{Fig_1}e we show the difference image for the two chiralities, demonstrating unidirectional propagation along the edge and little scatter into the bulk. 
For the longest evolution times ($t = 10\,\si{\milli\second}$), the center-of-mass position of the cloud travels approximately $17 a$, and the fastest $20\%$ of the atoms travel more than $30 a$. 
This underlines the potential of our experimental protocol for probing the topological properties of Floquet topological systems, where knowledge about the Chern numbers is insufficient, or where conventional methods for detecting bulk geometric properties are not applicable, e.g., in the presence of disorder.

\begin{figure*}[!htb]
\includegraphics{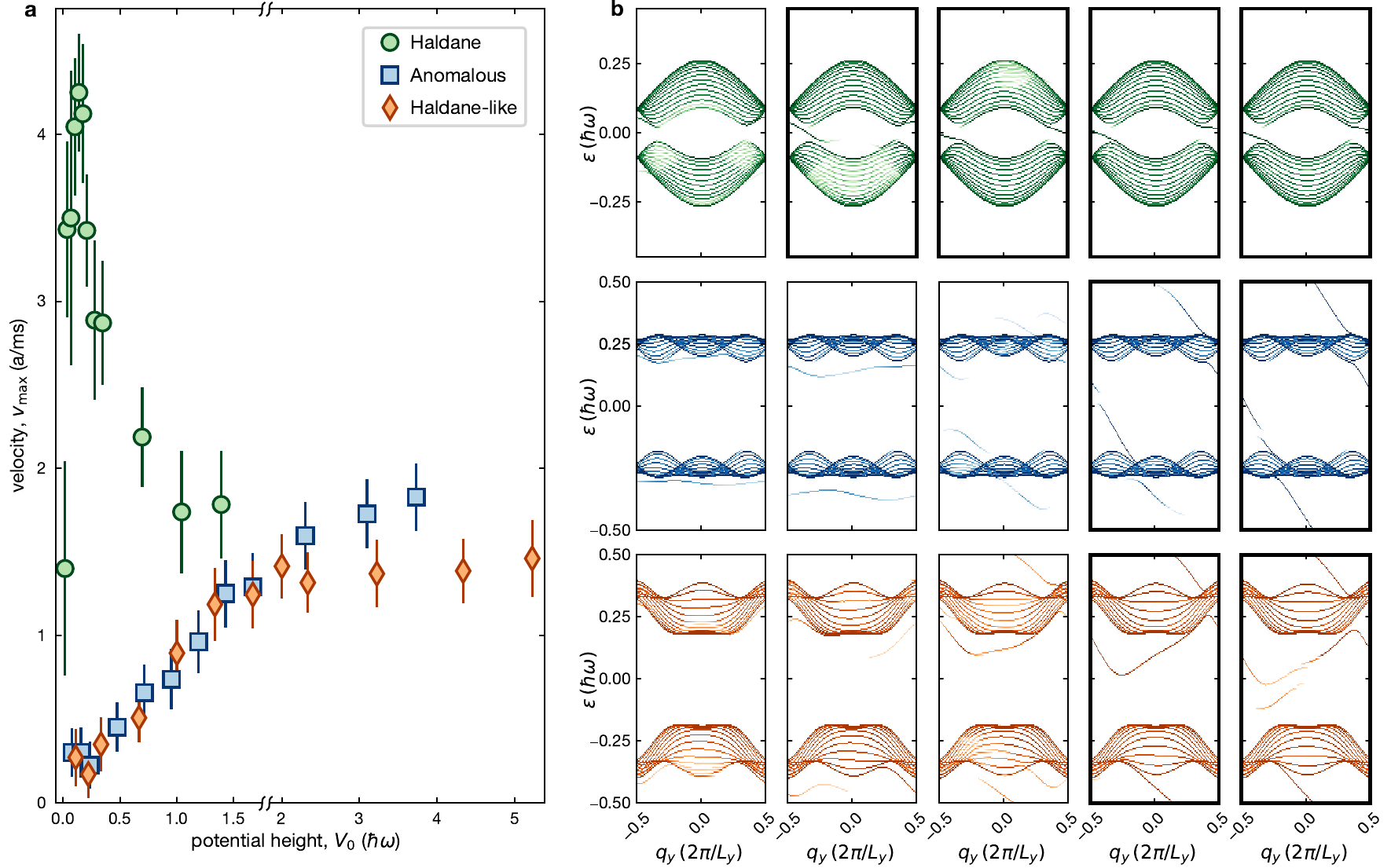}
\vspace{-0.cm}
\caption{\textbf{Emergence of edge states with increasing potential height $V_0$.}
\textbf{a}, Measured maximum edge state velocity $v_\text{max}$. 
The modulation parameters for the three distinct topological regimes are indicated in \figref{Fig_1}b.
Each data point is the average of three data sets that have been taken on different days, 
and the error bars are calculated from their standard deviation 
and the uncertainty due to the evaluation of the edge state velocity. 
\textbf{b}, Numerical simulations of the quasienergy spectrum 
using a step-wise modulated tight-binding model on a semi-infinite system 
as explained in the Methods for the three topological regimes: 
Haldane (top row), anomalous (middle row) and Haldane-like (bottom row).
In the finite direction, an infinitely sharp potential step of height $V_0$ is applied in the middle of the system. 
The spectra show the eigenenergies whose eigenstates have a significant overlap with this low-potential region (Methods). 
The potential height is increased from the left to the right with $V_0/(\hbar\omega) = \{0.05,\, 0.1,\, 0.5,\, 1.0,\, 2.0\}$.
The spectra where the edge modes are clearly visible are highlighted with a bold black frame.
}
\label{Fig_3}
\end{figure*}

While we observe a robust, large overlap of the initial wave packet in the anomalous regime, this is not the case in the Haldane regime~\cite{haldaneModelQuantumHall1988}. 
To achieve a good overlap of the initial wave packet with the edge modes, both the phase profile along the edge and the transverse extension of the initial wave packet have to match the one of the edge state~\cite{goldmanCreatingTopologicalInterfaces2016}. 
To gain more insight into the structure of the eigenmodes of the system, we perform numerical simulations with a step-wise modulated tight-binding model~\cite{kitagawaTopologicalCharacterizationPeriodically2010} where the nearest-neighbour tunnel couplings are amplitude-modulated in a chiral manner between a large and a small value, denoted as $J_1$ and $J_1^\prime$ respectively.
The numerical system is a semi-infinite strip with infinitely sharp walls along $x$ (zigzag termination) and periodic boundary conditions along $y$; the size of the unit cell is $L_y= \sqrt{3} a$ (see Methods for details).
In Fig.~\ref{Fig_2} we show the quasienergy spectrum in the first Floquet Brillouin zone for parameters in the anomalous (\figref{Fig_2}a) and Haldane regime (\figref{Fig_2}b).
To further understand the properties of the edge modes, we display exemplary wave functions in both regimes. 
We find that in the anomalous regime, the edge state in the $\pi$-gap does not exhibit any phase gradient along the edge, while in the Haldane regime the edge state exhibits a phase gradient of $\pi/L_y$ along the edge. 
Since the initial wave packet has a flat phase profile, it has a good overlap with the former, but not with the latter.

To optimize the overlap with the edge mode in the Haldane regime with regards to the spatial extent and the phase profile of the initial wave packet, we imprint a phase gradient by applying a kick with the tweezer (see Methods) and adjust the size by varying the radial trap frequency $\omega_\perp$.
In order to evaluate the fraction of atoms in the edge mode, we let the wave packet evolve for a time long enough to separate atoms near the edge from those scattered into the bulk modes (insets of \figref{Fig_2}c).
The fraction of atoms in the edge state is then evaluated (see Methods) as a function of the imprinted phase gradient.
For small $\omega_\perp$ (light green data points), we observe an overall poor overlap with the edge mode, which however increases as we increase the phase gradient close to the theoretical optimum of $0.58\pi/a$.
Decreasing the initial spatial extent of the cloud with a tighter tweezer further reduces scattering into the bulk (dark green data points).
In this regime, the phase gradient likely does not play a significant role, since primarily one site is significantly populated, ensuring a large overlap with the edge mode.

In the anomalous regime instead we find an extremely robust behavior, 
where the fraction of atoms in the edge mode is largely independent of the properties of the initial wave packet. 
In this regime, the initial wave packet is projected onto both edge modes, 
one in the $0$-gap and one in the $\pi$-gap, for each quasi-momentum (\figref{Fig_2}a). 
Varying the parameters of the initial state should only affect the relative weight between the two modes, 
which we cannot independently resolve in the experiment.

\begin{figure}[!htb]
\includegraphics{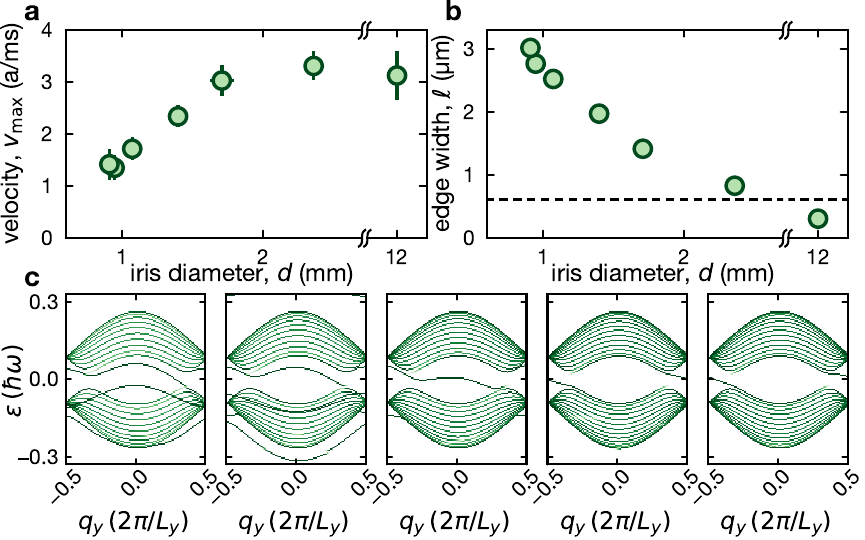}
\vspace{-0.cm} 
\caption{\textbf{Edge state velocity for varying edge width.}
\textbf{a}, Measured edge state velocity as the Fourier plane iris is closed for a repulsive potential with height $V_0/h=1.1\,\si{\kilo\hertz}$ in the Haldane regime (modulation parameters marked in \figref{Fig_1}b) for $J_0 = 1.1(1)\,\si{\kilo\hertz}$, $\omega_\perp=2\pi\times2.0\,\si{\kilo\hertz}$ and a phase gradient of $0.43(1)\pi/a$. The data points are averages of three individual datasets, and the error bars are evaluated from their standard deviation and the uncertainty of the evaluation of the velocity for each dataset.
\textbf{b}, Estimated edge width of the pattern as a function of the diameter $d$ of the iris in the Fourier plane. The dashed line corresponds to the theoretical resolution limit of the microscope objective. The edge width measured for an iris that is fully open (rightmost data point) is thus limited by the finite resolution of the objective. 
\textbf{c}, Floquet spectra simulated with the step-wise modulated tight-binding model in the Haldane regime (table in methods). The color indicates the overlap with a region covering the low-potential region. The height of the edge is fixed at $V_0/(\hbar\omega) = 1.44$, and its width is varied from left to right: $\ell/a = \{6, 4, 2, 1, 0.1\}$}
\label{Fig_4}
\end{figure}

In our experimental setup the topological interface is generated by a potential step.
In order to understand the characteristic energy scale of the potential needed for an edge mode to emerge at the interface, we investigate the maximum velocity $v_\text{max}$ of the edge state as a function of the height of the potential $V_0$ (Methods).
We study three distinct topological regimes as indicated by the colored hexagons in \figref{Fig_1}b:
the Haldane regime with $(W_0,W_\pi)=(1,0)$~\cite{haldaneModelQuantumHall1988}, which exhibits a conventional topological edge mode around zero energy and Chern numbers $\mathcal{C}^\pm=\mp1$;
the anomalous regime with $(W_0,W_\pi)=(1,1)$~\cite{kitagawaTopologicalCharacterizationPeriodically2010,winterspergerRealizationAnomalousFloquet2020}, which exhibits an additional anomalous edge mode in the $\pi$-gap with $\mathcal{C}^\pm=0$ 
and a third Haldane-like regime, with $(W_0,W_\pi)=(0,1)$~\cite{winterspergerRealizationAnomalousFloquet2020}, with Chern numbers $\mathcal{C}^\pm=\pm1$, but where the topological edge mode is located in the non-trivial $\pi$-gap. 

In all three topological regimes we find that the group velocity of the atoms 
starts to increase as we increase the height of the potential (\figref{Fig_3}a).
In the Haldane regime the maximum velocity is reached for a potential on the order of $\approx 0.1\hbar\omega$,
which matches the characteristic energy scale of the energy gap between the two bands. 
For larger potential depths the velocity starts to gradually decrease. 
Intuitively, one may expect a saturation of the velocity as soon as the edge mode is 
established at the interface. 
We attribute the gradual slowing down observed in our experiment to potential corrugations 
and a smaller slope of the potential edge near the bottom of the potential, 
which becomes more significant as we increase its height.
The general trend observed in the anomalous and Haldane-like regime is similar, 
but distinctly different from the Haldane regime in terms of absolute values.
Here, we find that the characteristic energy scale for the potential needed to reach the maximum group velocity
is on the order of $\approx 2\hbar\omega$. For larger values we observe a saturation behavior.

To support the experimental results, we numerically investigate a semi-infinite strip geometry, 
where the potential energy of one half of the strip in the finite direction is increased by $V_0$.
In order to visualize the appearance of the topological interface, 
we show the eigenenergies in the low-potential region in \figref{Fig_3}b (see Methods).
Similar to our experimental results we find that in the Haldane regime, 
the edge state in the $0$-gap emerges at a characteristic energy scale given by the size of the $0$-gap, 
which is about an order of magnitude smaller than $\hbar\omega$.
For the same potential height there are no clear signatures of edge modes 
in the anomalous and the Haldane-like regimes in any of the two gaps. 
Instead we find a characteristic energy scale on the order of $\hbar\omega$ for them to appear. 
In \figref{Fig_3}b we highlight all spectra where we believe hat edge modes can clearly be identified.
This behavior is qualitatively consistent with our observations.

The finite width of the potential edge has a large impact on the group velocity of the particles in the edge modes, 
as the dispersion of the edge mode hybridizes with the bulk modes, resulting in a significant reduction of the velocity~\cite{stanescuTopologicalStatesTwodimensional2010, buchholdEffectsSmoothBoundaries2012, goldmanDirectImagingTopological2013}.
We investigate this behavior by tuning the width of the potential edge in the Haldane regime (\figref{Fig_4}a).
The width of the edge is controlled by varying the diameter $d$ 
of an iris placed in the Fourier plane of the imaging system 
that is used to project the DMD potential into the atomic plane (Methods).
Because of the incoherent illumination closing the iris leads to a reduction of the potential height. 
This is compensated by an increase of the total power of the beam 
to ensure the same potential height $V_0$ for all measurements.
The width of the edge is measured by imaging the pattern in an intermediate plane, 
fitting the edge profile with an error function and extracting the characteristic length of this fit~\cite{DefinedLength92}.
This length is then multiplied by the magnification between the intermediary plane and the atoms, 
which was calibrated independently. 
Figure~\ref{Fig_4}b shows the resulting width as a function of the iris diameter $d$.
Note that the actual experimental value of the edge width at the atom position 
is most likely further increased by imperfect alignment and residual aberrations.
The dashed line in \figref{Fig_4}b illustrates the theoretical resolution limit.
We find that a smoothened edge leads to a significant reduction of the edge state velocity, as expected~\cite{stanescuTopologicalStatesTwodimensional2010, buchholdEffectsSmoothBoundaries2012, goldmanDirectImagingTopological2013, gebertLocalChernMarker2020}. This is further confirmed by numerical simulations using the simple step-wise modulated tight-binding model (\figref{Fig_4}c).

In conclusion, we have demonstrated an experimental protocol for the preparation
and manipulation of topological edge modes with ultracold atoms in real space. 
We have presented a detailed study on the preparation of atoms in the edge modes 
and how the effectiveness of the protocol depends on the initial-state parameters.
Making use of the unique control over the shape of the potential, we further investigated how the 
edge modes emerge at the topological interface as the height of the potential and its sharpness is increased.
Our study provides an essential new tool to probe the topological features of different phases of matter 
with ultracold atoms, in particular in slowly-driven systems and in the presence of disorder, where other techniques are not applicable. 
We anticipate, that his will enable experimental investigations of the rich phases that can arise in non-interacting disordered systems~\cite{liTopologicalAndersonInsulator2009, titumDisorderInducedFloquetTopological2015, stutzerPhotonicTopologicalAnderson2018} or interacting ones~\cite{titumAnomalousFloquetAndersonInsulator2016, nathanAnomalousFloquetInsulators2019}.
It also bridges the gap with techniques routinely used in solid-states and photonics systems~\cite{vanhoutenQuantumPointContacts1996, jiElectronicMachZehnder2003, tambascoQuantumInterferenceTopological2018, carregaAnyonsQuantumHall2021} to study the coherence and transport properties of these edge states~\cite{buttikerEdgeStatePhysicsMagnetic2009}.

\paragraph*{\textbf{Note:}} During completion of this manuscript, we became aware of recent theoretical studies, where wave packet dynamics in Floquet-driven topological lattices has been studied~\cite{martinezWavePacketDynamics2023}.


\paragraph*{\textbf{Acknowledgements}}
We thank N.~\"Unal, M.~Martínez, N.~Goldman and M.~Di Liberto for fruitful discussions. We further thank A.~Alberti for the lattice laser. This work was funded by the Deutsche Forschungsgemeinschaft (DFG, German Research Foundation) via Research Unit FOR 2414 under project number 277974659. The work was further supported under Germany’s Excellence Strategy -- EXC-2111-- 3908148. J.A. was funded by the Alfried
Krupp von Bohlen und Halbach foundation. R.S.-J. has received funding from the European Union’s Horizon 2020 research and innovation programme under the Marie Sklodowska-Curie grant agreement 101028339.

%
\paragraph*{\textbf{Data availability}}
The data that support the plots within this paper and other findings of this study are available from the corresponding author upon reasonable request.

\paragraph*{\textbf{Code availability}}
The code that supports the plots within this paper are available from the corresponding author upon reasonable request.


\section*{Methods}
The Methods section contains additional information about the expansion of the initial wave packet in the static lattice (Sec.~\ref{sec:expansion}), 
calibration measurements for imprinting the phase gradient (Sec.~\ref{sec:kick_calibration}),
a description of how the population of atoms in the edge mode is evaluated (Sec.~\ref{sec:edge_fraction}),
details on the numerical simulations using a semi-infinite strip (Sec.~\ref{sec:numerical_simulations}),
an exemplary evaluation of the edge-state velocity (Sec.~\ref{sec:velocity_measurement}),
a calibration of the height of the potential edge (Sec.~\ref{sec:edge_calibration})
and a description of the optical setup for the implementation of the programmable optical potential (Sec.~\ref{sec:DMD_setup}).
Additional information about the tweezer setup, calibration and lifetime measurements
is given in the Supplementary Material.


\begin{figure}[!ht]
\includegraphics{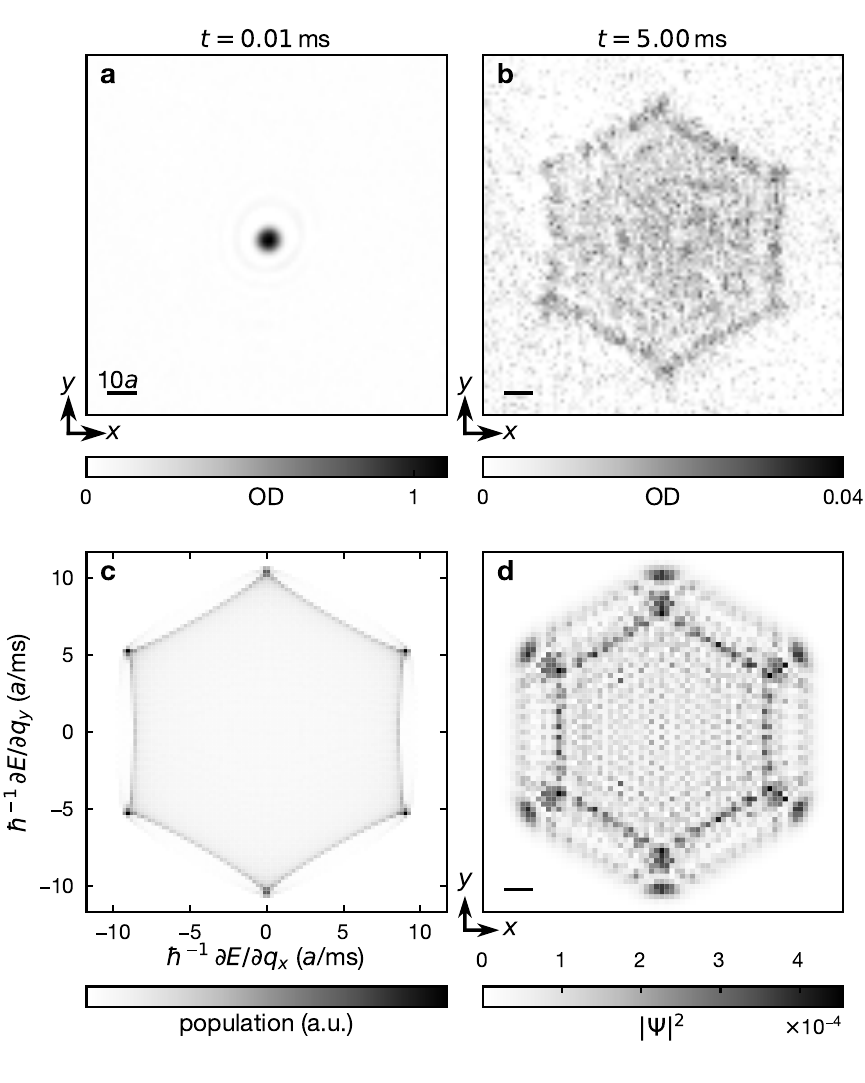}
\vspace{-0.cm} \caption{\textbf{Evolution of the initial wave packet in the bulk of a static lattice.} 
\textbf{a}, Spatial distribution of the initial wave packet $10\,\mu$s after release into the lattice. The wavepacket is prepared at $\omega_\perp/(2\pi) = 1.3(1)\,\si{\kilo\hertz}$.
\textbf{b}, Real-space distribution of the atoms in the static lattice after an expansion time of $5\,\si{\milli\second}$.
The data shown in \textbf{a} and \textbf{b} are an average over 302 individual realizations.
\textbf{c}, Group velocity distribution for a uniformly filled lowest band of the honeycomb lattice.
\textbf{d}, Numerical simulation of the expansion dynamics in a tight-binding model starting from a state occupying a single site. 
The resulting probability distribution $|\psi(\mathbf{r},t)|^2$ is binned to match the number of pixels in \textbf{b}, the scale bar corresponds to $10a$.
}
\label{Fig_SM10}
\end{figure}

\subsection{Expansion dynamics of the initial wave packet in a static lattice}
\label{sec:expansion}
Figure~\ref{Fig_SM10}a shows the initial density distribution of the atoms $10\,\si{\micro\second}$ 
after releasing them from the tweezer into the static honeycomb lattice. 
The measured signal is the convolution of the actual density distribution in the trap 
with the point-spread function of the imaging system (NA=$0.5$).
To demonstrate that our initial wave packet evolves coherently in the lattice and occupies sufficiently 
high quasi-momentum states in the Brillouin zone, we observe its expansion in the bulk 
of a static lattice without any repulsive edge potential (\figref{Fig_SM10}b).

To this end, the tweezer trap is switched off abruptly and the evolution of the
initial wave packet in the static lattice is monitored.
After a short evolution time we observe a characteristic hexagonal shape,
which is expected to emerge during coherent expansion dynamics for a homogeneously populated band~\cite{schneider_fermionic_2012,ronzheimer_expansion_2013}.
This is explained by the quasi-momentum dependent group velocities $\partial_{\mathbf{q}} E(\mathbf{q})/\hbar$, 
as determined by the dispersion relation of the energy bands in the honeycomb lattice (\figref{Fig_SM10}c).
In contrast, for an incoherent random walk  
a Gaussian distribution would be measured~\cite{durQuantumWalksOptical2002}, 
and if only the center of the Brillouin zone was populated, 
only the quadratic dispersion relation of the lowest band would be probed, 
masking the edges of the hexagonal shape.

Additionally, due to the interference of all the independent quasi-momenta, 
a coherent evolution displays an interference pattern within the hexagonal shape.
In \figref{Fig_SM10}d we show numerical simulations of an initial state that is fully localized 
on a single lattice site after $5\,\si{\milli\second}$ of expansion.
The simulations are performed with a two-band model whose parameters are close to the experimental realization.
These measurements confirm a sufficient localization of the initial wave packet and 
support a coherent evolution in the lattice.
We further confirmed that we observe a similar quality of the expansion dynamics 
when letting the atoms evolve in the bulk of the modulated lattice in the three regimes.


\begin{figure}[!t]
\includegraphics{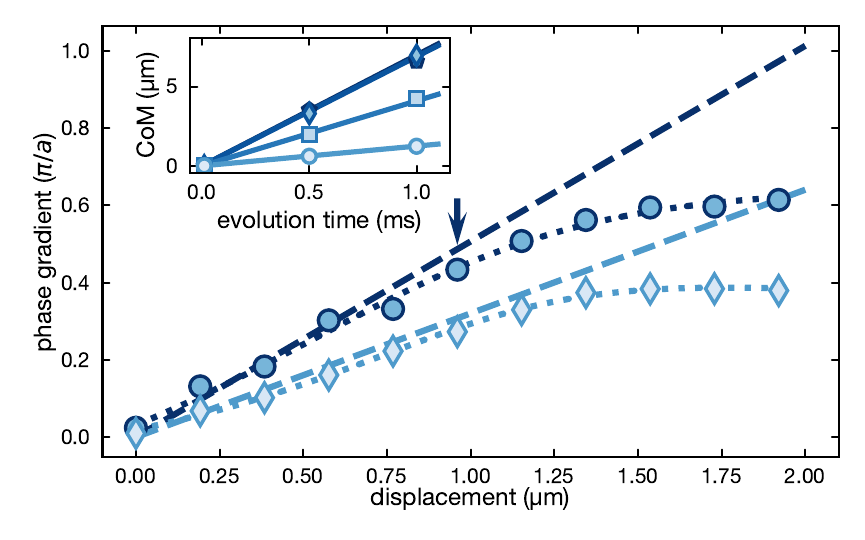}
\vspace{-0.cm} 
\caption{\textbf{Phase gradient induced by the kick.}
The velocity of the wave packet is measured as a function of the total displacement of the tweezer for two different in-plane trapping frequencies: $\omega_\perp/(2\pi) = 1.3(1)\,\si{\kilo\hertz}$ in light blue, and $2.0(1)\,\si{\kilo\hertz}$ in dark blue.
The dashed lines are the analytic predictions from the motion of a cloud in a harmonic potential, and the dotted lines are the interpolation of the measurements with a polynomial of degree five.
The inset shows the averaged measurement of center of mass (CoM) of the atoms after the kick and a variable evolution time in the optical dipole trap for the lowest value of $\omega_\perp$ and displacements of $0.19\,\si{\micro\meter}$ (circles), $0.78\,\si{\micro\meter}$ (squares), $1.36\,\si{\micro\meter}$ (diamonds), and $1.94\,\si{\micro\meter}$ (pentagons). The depicted datapoints are an average of 6 individual realizations.
The inset also shows the linear weighted fits from which the velocity is extracted.
The associated uncertainty of the slope is used as the error bar for the main graph, and is smaller than the size of the markers. 
The arrow at displacement $\approx 1\,\si{\micro\meter}$ indicates the phase gradient that was used in all measurements to populate the edge mode in the Haldane regime.}
\label{Fig_SM2}
\end{figure}

\subsection{Initial velocity kick calibration}
\label{sec:kick_calibration}
A phase gradient is provided to the initial cloud by giving it a kick.
To this end the experimental sequence is slightly modified.
Shortly before switching off the tweezer, it is displaced parallel to the edge to give an initial velocity to the wave packet, thus imprinting a phase gradient.
The velocity kick is given by applying a linear ramp to the radio frequency sent to the AOD, resulting in a linear displacement of the tweezer in the $y$ direction during a time $\delta_t$: $y^\mathrm{tw}(t) = y^\mathrm{tw}_\mathrm{i} + (y^\mathrm{tw}_\mathrm{f} - y^\mathrm{tw}_\mathrm{i}) t/\delta_t$.
At the end of the linear ramp, the tweezer is abruptly switched off.
Classically, the center of mass of the atomic cloud $\langle y\rangle(t)$ obeys the following equations:
\begin{equation}
m_\text{K} \frac{\mathrm{d}^2\langle y \rangle}{\mathrm{d}t^2}= -\nabla U(y,t),
\end{equation}
where $U(y,t)$ is the optical potential of the tweezer, approximated by a parabola:
\begin{equation}
U(y,t) = \frac{m_\text{K} \omega_\perp^2}{2}\left(y - y^\mathrm{tw}(t)\right)^2.
\end{equation}
The equation of motion can be integrated with the initial conditions $\frac{\mathrm{d}\langle y\rangle}{\mathrm{d}t}(0) = 0$ and $\langle y\rangle(0) = y_\mathrm{i}$:
\begin{equation}
 \langle y\rangle(t) = y^\mathrm{tw}_\mathrm{i} +  \frac{y^\mathrm{tw}_\mathrm{f} - y^\mathrm{tw}_\mathrm{i}}{\omega_\perp \delta_t}\left[\omega_\perp t - \sin(\omega_\perp t)\right],
\end{equation}
which indicates that at the end of the ramp $t = \delta_t$, the final velocity of the center of mass of the atoms $v_y$ is:
\begin{equation}
v_y(t=\delta_t) = \frac{y^\mathrm{tw}_\mathrm{f} - y^\mathrm{tw}_\mathrm{i}}{\delta_t}\left[1 - \cos(\omega_\perp \delta_t)\right].
\end{equation}
For a given tweezer frequency $\omega_\perp$, the final velocity can be maximized: 
the function $x \mapsto \left[1-\cos(x)\right]/x$ reaches its maximum for $x\approx 2.33$, therefore one can choose $\delta_t = 2.33/\omega_\perp$. The amplitude of the displacement $y^\mathrm{tw}_\mathrm{f} - y^\mathrm{tw}_\mathrm{i}$ can then be chosen independently, provided that it does not exceed the scale within which the parabolic assumption is valid.

We experimentally measure the final velocity imposed by the kick by letting the atoms evolve in the dipole trap after applying the linear ramp and subsequently switching off the tweezer and the optical lattice.
We measure the position of the center of mass of the cloud as a function of the evolution time that is varied between $0$ and $1\,\si{\milli\second}$.
The position in the direction of the kick is linear with time, and we extract the corresponding velocity with a linear fit (inset of \figref{Fig_SM2}).
The measured velocities range between $0$ and $10\,\si{\micro\meter/\milli\second}$, and are converted into a phase gradient via
\begin{equation}
\nabla \phi = \frac{m_\text{K}}{\hbar} v_y,
\end{equation}
as shown in \figref{Fig_SM2} as a function of the displacement of the tweezer for two values of $\omega_\perp$.
The dashed lines correspond to the theoretical prediction derived above, which is in very good agreement with the measurements for small displacements.
At larger displacements, the measurements depart from this prediction due to the anharmonicity of the tweezer, whose waist is on the order of $1\,\si{\micro\meter}$.
As indicated by the vertical scale of the graph, this method allows to prepare phase gradients large enough to engineer a significant phase shift between two neighbouring sites.
The value of the phase gradient reported in \figref{Fig_2} of the main text is obtained by interpolating the values  measured here with a polynomial of order five, shown here as the dotted lines.
Finally, the parameters used to prepare the cloud with a non-zero phase gradient to observe edge states -- in particular in the Haldane phase -- is indicated by the blue arrow on \figref{Fig_SM2}.


\begin{figure}[!t]
\includegraphics{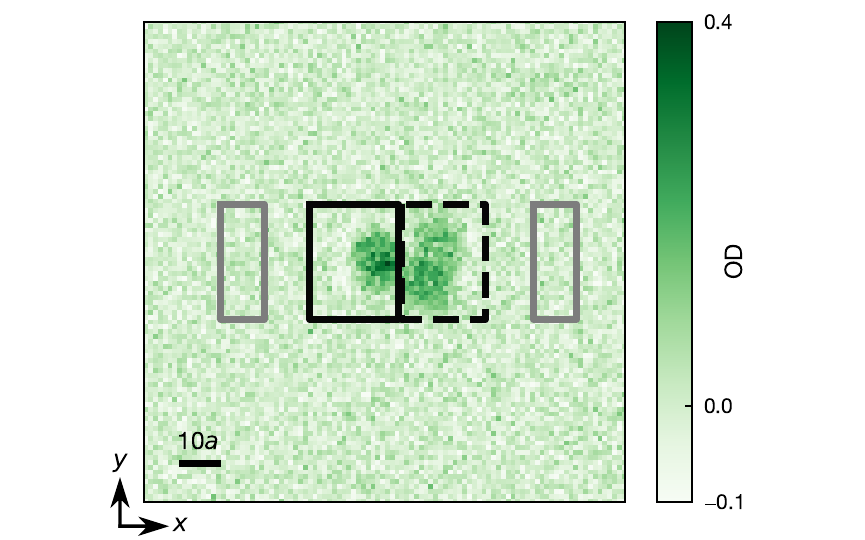}
\vspace{-0.cm}
\caption{\textbf{Evaluation of the fraction of atoms populating the edge mode.}
Averaged absorption image after $1.5\,\si{\milli\second}$ of time evolution, 
together with the respective regions of interest used for the evaluation: 
close to the edge (solid black lines) and in the bulk (dashed black lines).
The two background regions of half the size are indicated by the gray lines. 
This example corresponds to the $\omega_\perp/(2\pi)=1.3(1)\,\si{\kilo\hertz}$ 
data point of \figref{Fig_2} shown in the bottom right inset.
}
\label{Fig_SM7}
\end{figure}

\subsection{Evaluation of the population fraction in the edge mode}
\label{sec:edge_fraction}
The edge population fraction, displayed in \figref{Fig_2}c, is determined by integrating the optical density (OD) in the respective regions of interest (\figref{Fig_SM7}).
We prepare the initial state with varying trap frequency $\omega_\perp/(2\pi)=1.3(1)\,\si{\kilo\hertz}$ and $2.0(1)\,\si{\kilo\hertz}$ of the optical tweezer and different initial phase gradients.
In order to ensure the optimum spatial overlap of the initial state with the edge mode, we additionally vary the position with respect to the edge potential.
The fraction of atoms loaded in the edge mode is evaluated after $1.5\,\si{\milli\second}$ evolution in the modulated lattice, when the atoms in the bulk and on the edge have spatially separated, ensuring that the two signals can be distinguished.
Several absorption images are taken with the same experimental parameters and averaged.
To determine the size of the region of interest, we verify that the integrated optical density is unaltered for small changes in the size of the region of interest.
Additionally, two regions of half the size of the previous ones are defined separated from the central region to evaluate the background value of the image (\figref{Fig_SM7}).
We independently sum the optical density of the edge and bulk regions of interest and subtract the summed optical density in the background regions to obtain the signal from the bulk $S_\mathrm{bulk}$ and edge $S_\mathrm{edge}$. These values are then used to compute the fraction $p_\mathrm{edge}$ of atoms in the edge region: 
\begin{equation}
p_\mathrm{edge} = \frac{S_\mathrm{edge}}{S_\mathrm{edge} + S_\mathrm{bulk}}.
\end{equation}
The error bar is evaluated using the standard deviation of the two values of the fraction obtained for the two opposite chiralities of the lattice modulation and the noise of the imaging system that is obtained from the standard deviation of the optical density in the background region.


\begin{figure}[!t]
\includegraphics{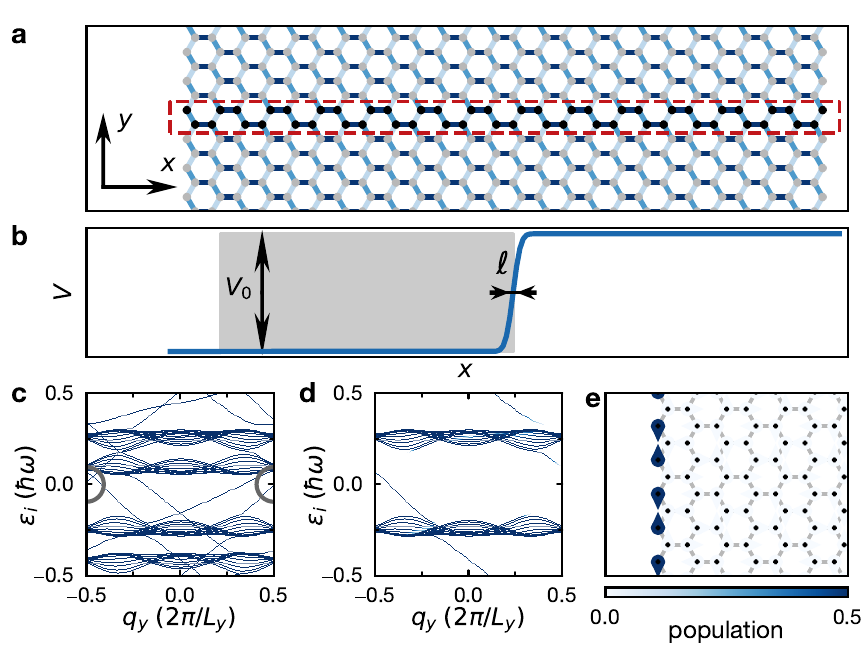}
\vspace{-0.cm} \caption{\textbf{Numerical simulations.}
\textbf{a}, Geometry of the lattice. The system consists of the sites located in the dashed red rectangle. In the $y$-direction the system is periodic, as indicated by the repetition of the system in pale colours. The three types of links between sites are indicated by the three shades of blue. The links that are weighted by a complex phase are those that cross the dashed lines.
\textbf{b}, Potential energy applied in the $x$-direction. The position of the edge is chosen in the center of the system, and its width $\ell$ and its height $V_0$ are free parameters. In this example, $\ell = a$. The shaded region is an example of a zone that can be used to selectively filter the spectrum.
\textbf{c}, Full quasienergy spectrum of the system with $J_1^\prime/J_1 = 0.1$, $\hbar\omega = 2\pi\times 1.5 J_1$, $\ell = a$ and $V_0 = J_1/2$. The system is in the anomalous regime, i.e., two pairs of edge states are visible, linking the two bulk energy bands in the two gaps of the model. The  gray circle indicates the eigenstate whose wavefunction is represented in \textbf{e}.
\textbf{d}, Exemplary quasienergy spectrum. For this example, the same spectrum as in \textbf{c} is plotted with a color scale that represents the overlap with the shaded region of graph \textbf{b}. 
\textbf{e}, Wavefunction of the eigenstate at $q_y = \pm \pi/L_y$ and at energy $\varepsilon=0$ corresponding to a numerical edge state. The position of the sites is represented as the black dots and the tunnelling bonds are shown in gray. Around these dots, the color indicates the modulus square of the wavefunction, and the direction of the arrow indicates its phase.}
\label{Fig_SM4}
\end{figure}

\subsection{Numerical simulations of the tight-binding model on a semi-infinite strip}
\label{sec:numerical_simulations}
For the numerical simulations we implement a time-dependent tight-binding model~\cite{kitagawaTopologicalCharacterizationPeriodically2010} on a hexagonal lattice with a finite dimension in $x$ and an infinite dimension in $y$. 
The system is represented in the dashed red rectangle in \figref{Fig_SM4}a: in the $x$-direction a system of length $L_x$ is implemented, and in the $y$-direction, the minimal length $L_y$ necessary to create a unit cell is generated.
The hexagonal lattice is oriented such that the edges of the system at $x = \pm L_x/2$ are zigzag edges, similarly to the experiment. 
This leads to $L_y = \sqrt{3}a$, and the size $L_x$ is chosen equal to $20a$, such that the edge effects cover a region in $x$ smaller that $L_x$, while keeping the total number of sites in the unit cell $N_s = 52$ relatively small to ensure fast numerical calculations.
In this lattice only nearest-neighbour hoppings are considered, and three types of links are distinguished depending on their orientation (see the three colours in \figref{Fig_SM4}).

\begin{table}[h]
\label{tab:1}
\centering
\begin{tabular}{c|c|c|c}
\textup{Variables}	& Haldane 	& Anomalous & Haldane-like 	\\
\hline
$J_1^\prime / J_1$  & 0.1 		& 0.1		& 0.1      	\\
$\hbar \omega / J_1$& $9\pi$  	& $3\pi$	& $2\pi$	\\
\hline
\end{tabular}
\caption{Variables used in the numerical simulations}
\end{table}

The time-periodic Hamiltonian with period $T = 2\pi/\omega$ is implemented by dividing each drive period into three steps of equal duration described by a Hamiltonian $\hat{H}_i$ ($i=\{1,2,3\}$).
During each of them, one of the three links has a large hopping amplitude $J_1$, while the two others have a low hopping amplitude $J_1^\prime$.
The hopping amplitude from site $i$ to site $j$ is associated with an additional phase factor $\E^{\pm \I q_y L_y}$  if the corresponding hopping links two neighboring unit cells, where the sign depends on the direction of the link along $y$; $q_y$ denotes the quasi-momentum along $y$.
This leads to three Hamiltonians that can be expressed in the form of $N_s\times N_s$ matrices with entries that depend on the quasi-momentum.
The order in which the links are switched from a low to a high amplitude during one period defines the two chiralities of the modulation.
Finally, a potential offset can be applied on each of the sites, which is described by the Hamiltonian $\hat{V}$.

The evolution operator $\hat{U}(T)$ during one period is then calculated as a function of $q_y$:
\begin{equation}
\hat{U}(T) = \E^{-\I\left(\hat{H}_3 + \hat{V}\right)\frac{T}{3\hbar}}\E^{-\I\left(\hat{H}_2 + \hat{V}\right)\frac{T}{3\hbar}}\E^{-\I\left(\hat{H}_1 + \hat{V}\right)\frac{T}{3\hbar}}.
\end{equation}
This operator is diagonalized to obtain its eigenvalues $\lambda_\mu$ and eigenstates $\ket{\psi_\mu}$, for $\mu = \{1,\ldots, N_s\}$. The associated quasi-energies $\varepsilon_\mu$ are then calculated as~\cite{rudnerAnomalousEdgeStates2013}:
\begin{equation}
\varepsilon_\mu = \frac{\hbar\omega}{2\pi\I} \ln(\lambda_\mu).
\end{equation}
The potential energy is modelled as
\begin{equation}
V(x) = \frac{V_0}{2}\left[\mathrm{erf}\left(\frac{2(x_0 - x)}{\ell}\right) + 1\right],
\label{eq:edge_potential}
\end{equation}
where $\mathrm{erf}(x)$ is the error function, $x_0$ is the position of the edge, chosen here in the middle of the system, and $\ell$ encodes the width of the edge~\cite{DefinedLength92}, as illustrated in \figref{Fig_SM4}b.
The spectrum $\{\varepsilon_\mu(q_y)\}$ is plotted as a function of the quasi-momentum $q_y$ in \figref{Fig_SM4}c, where the simulation is performed with a value of $\omega$ and $V_0$ such that the system is in the anomalous regime and the potential $V_0$ is large enough to create an edge state in the center of the system.
We find that each of the two bands of the model is duplicated: the first one represents the left part of the system, and the second one the right part of the system, shifted by $V_0$ in energy, with respect to the first one.

Additionally, if a region $\mathcal{S}$ is defined on the lattice, the projection $\mathcal{P}(\mu,\mathcal{S})$ of each of the eigenstates on this region can be calculated according to:
\begin{equation}
\mathcal{P}(\mu,\mathcal{S})  = \sum_{\mathrm{site}\, j}\mathbbm{1}_{\mathcal{S}}(j) \braket{j|\psi_\mu},
\end{equation}
where $\mathbbm{1}_{\mathcal{S}}(j) = 1$ if site $j$ belongs to the region $\mathcal{S}$ and 0 otherwise.
For example, the shaded area of \figref{Fig_SM4}b is chosen to filter the spectrum of \figref{Fig_SM4}c.
The same spectrum is shown in \figref{Fig_SM4}d, where each point has a color that represents the projection $\mathcal{P}(\mu,\mathcal{S})$.
For clarity the points with an overlap smaller than 1/20th of the maximal overlap are not shown.
In this figure one can see that the bands corresponding to the high part of the potential energy are removed, as well as the edge states on the (numerical) left edge.
Only the edge state at the potential edge in the center of the system remains.

Finally, the wavefunction of any of the eigenstates can be retrieved and displayed. In Fig.~\ref{Fig_SM4}e we show the wavefunction of the (numerical) edge state at $q_y = -\pi/L_y$ and at energy $\varepsilon=0$, where the information about the amplitude (color shading) and phase (arrow orientation) of the wavefunction is indicated. This is the edge mode of the $0$-gap in the anomalous phase, which has the same phase gradient along the edge as the edge mode of the Haldane phase.


\begin{figure}[!t]
\includegraphics{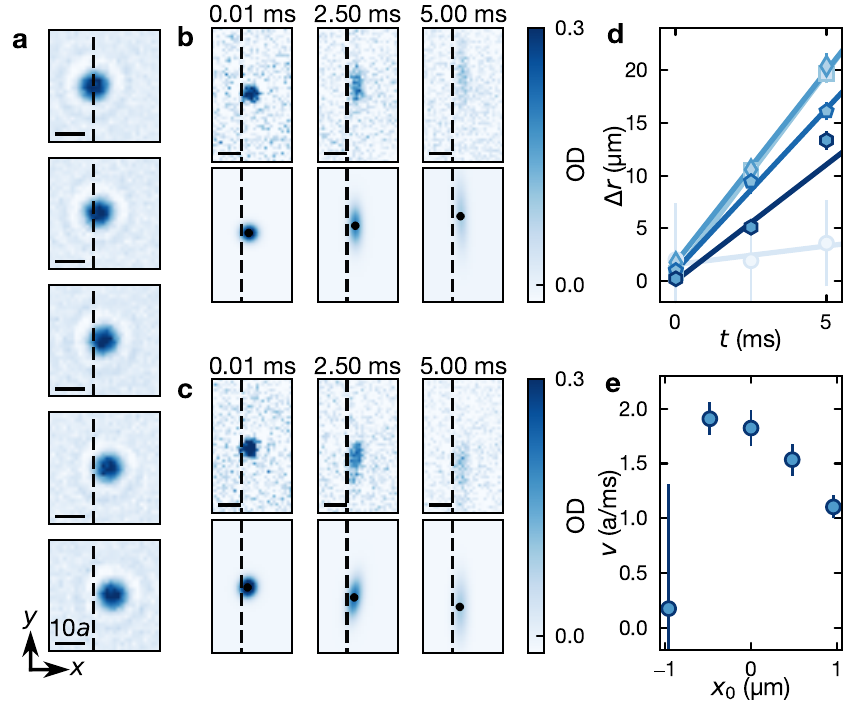}
\vspace{-0.cm}
\caption{\textbf{Determination of the edge state velocity.}
\textbf{a}, Insitu images of atoms in the optical tweezer. 
The position of the tweezer is varied from top to bottom by steps of $0.48\,\si{\micro\meter}$ in the $x$ direction (perpendicular to the orientation of the edge), as emphasized by the vertical dashed line.
The pictures are an average of five individual experimental realizations, performed without the edge potential in order to better see the displacement of the tweezer. The scale bar corresponds to $10a$.
\textbf{b-c}, Evolution of the cloud for the two chiralities. The first line shows the average of the absorption images after an increasing evolution time, displayed with the same colorscale.
The second line shows the result of the Gaussian fit that is performed on the averaged images.
On these fits, the center of the Gaussian is indicated as a black dot.
The error bar, which stems from a bootstrap analysis, is smaller than the marker.
\textbf{d}, The absolute distance $\Delta r$ between the center-of-mass positions of the time-evolved clouds with the two different chiralities is plotted as a function of the time.
The three different markers correspond to the three initial positions, and the corresponding solid lines are linear fits.
\textbf{e}, The slope of the fit is divided by a factor of two to obtain the average velocity of the edge state.
The measured velocity is plotted as a function of the initial position, and the final value that is selected is the maximum value of these points.
For the leftmost tweezer position the error bar is very large: the atoms are released on top of the potential step and no reliable velocity can be extracted.}
\label{Fig_SM3}
\end{figure}

\subsection{Velocity measurement}
\label{sec:velocity_measurement}
We measure the velocity of the edge state by maximizing the spatial overlap of the initial cloud and the edge mode.
To this end, the initial position of the tweezer is varied with respect to the edge by steps of approximately $0.5\,\si{\micro\meter}$ (\figref{Fig_SM3}a).
We then release the cloud and observe its subsequent evolution for various durations, and for the two opposite chiralities of the modulation scheme.
The top rows of figures~\ref{Fig_SM3}b and c show the evolution for the two chiralities in the anomalous regime and with an edge height of $V_0=h\times 19\,\si{\kilo\hertz}$, where each picture is the average of $N_\mathrm{im}$ experimental images.
The position of the center of mass of these observed clouds are evaluated 
by fitting the averaged images with a Gaussian function with its center, its amplitude, 
its two sizes, its offset and the orientation of its eigenaxes as free parameters.
The error bars of these fitted parameters are estimated with a bootstrap method:
Among the $N_\mathrm{im}$ experimental images, a random draw with replacement of $N_\mathrm{im}$ of these images is performed.
The chosen images (with possible repetition) are averaged and the resulting image is fitted with the same Gaussian function.
This random drawing, averaging and fitting procedure is repeated $20$ times, thus providing as many estimates for the parameters of the Gaussian.
The error bar for the fitted parameters is given by the standard deviation of these obtained values.

Figure~\ref{Fig_SM3}d shows the distance $\Delta r$ between the center of mass of the wavepacket for the two chiralities plotted as a function of the time of evolution.
The separation of the two clouds is linear with time: with one chirality of the modulation the cloud moves with average velocity $+v$ in the $y$-direction, and for the other chirality it moves with velocity $-v$.
The slope of $\Delta r$ as a function of evolution time is extracted by a linear fit, and is divided by a factor of two to obtain the average velocity of the edge state. The error is estimated by taking the error of the linear fit.
This velocity is extracted for all the initial positions of the tweezer with respect to the edge, as shown in \figref{Fig_SM3}e, and displays a maximum when the overlap of the initial wave packet with the edge mode is maximized.
The observable that is reported in the main text is thus the maximal velocity that has been measured, along with its error bar.


\begin{figure}[!t]
\includegraphics{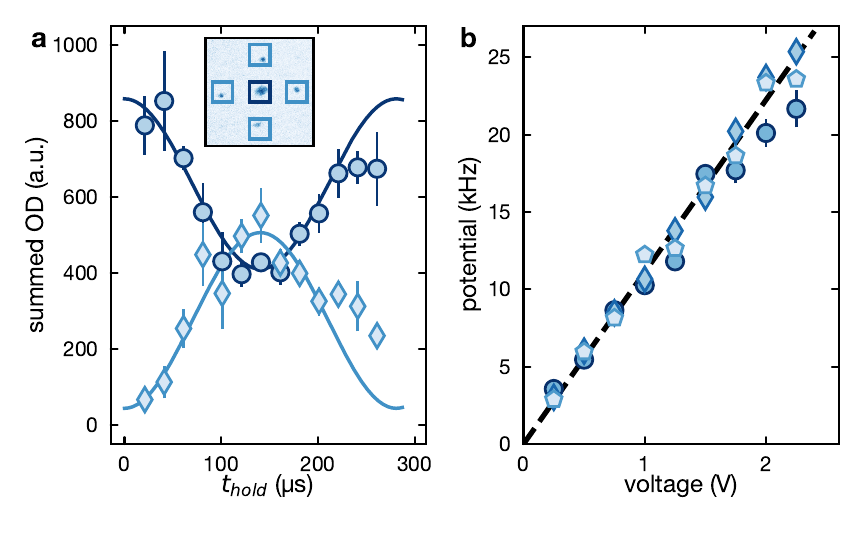}
\vspace{-0.cm}
\caption{\textbf{Calibration of the height of the repulsive potential.}
\textbf{a}, Measurement of the summed optical density in the regions corresponding to the zeroth (circles) and the first (diamonds) diffraction orders.
The error bars correspond to the standard deviation of the population in the respective order for images taken with the same parameters.
The solid lines show the sinusoidal fits from which the frequency is extracted.
The inset shows an exemplary image, where the different regions of interest for evaluating the optical density are indicated (light blue box: first diffraction order; dark blue box: zeroth order).
\textbf{b}, Height of the repulsive potential evaluated using sinusodial fits as shown in \textbf{a} as a function of the voltage used to control the laser intensity for three values of $d_\mathrm{sq}$: $1.0\,\si{\micro\meter}$ (circles), $1.3\,\si{\micro\meter}$ (diamonds) and $1.7\,\si{\micro\meter}$ (pentagons).
The error bars correspond to the uncertainty of the fit, and the dashed black line is a weighted linear fit of all the measured points, which has a slope of $11.12(3)\,\si{\kilo\hertz/\volt}$.}
\label{Fig_SM1}
\end{figure}

\subsection{Edge height calibration}
\label{sec:edge_calibration}
In order to calibrate the height of the repulsive potential $V_0$, we abruptly switch on a chequerboard pattern on the DMD in order to study the diffraction of a large BEC as a function of the hold time $t_\mathrm{hold}$ during which the DMD beam is on, for a given beam intensity.
The period of the chequerboard pattern is denoted as $2 d_\mathrm{sq}$. We detect the diffracted atoms after an evolution time of $10\;\si{\milli\second}$ in the optical dipole trap (see inset of \figref{Fig_SM1}a), and measure the population in the zeroth and first diffraction orders.
The first-order diffraction orders appear at positions corresponding to the wave-vector $k_1\propto 1/d_\mathrm{sq}$, associated with the kinetic energy $E_\mathrm{R} = \hbar^2k_1^2/(2m_K)$.
In our experiments, this kinetic energy does not exceed $h\times 1\,\si{\kilo\hertz}$.
For hold durations $t_\mathrm{hold}\approx V_0/\hbar$, and for $V_0>E_\mathrm{R}$, the diffraction experiments are in the so-called Raman-Nath regime~\cite{morschDynamicsBoseEinsteinCondensates2006}.
In this regime, the populations $p_0$ and $p_{\pm 1}$ in the observed orders are:
\begin{align}
p_0(t_\mathrm{hold}) &\propto \cos^2\left(\frac{V_0 t_\mathrm{hold}}{2h}\right),\\
p_{\pm 1}(t_\mathrm{hold}) &\propto \sin^2\left(\frac{V_0 t_\mathrm{hold}}{2h}\right),
\end{align}
where $V_0$ is the height of the potential on the atoms.
We fit these populations as a function of $t_\mathrm{hold}$ with these functions to extract $V_0$ (\figref{Fig_SM1}a).
This procedure is performed for various values of the potential height $V_0$ by varying the laser intensity, and for three values of $d_\mathrm{sq}$ to check the consistency of the method.
The value of $V_0$ is presented in \figref{Fig_SM1}b as a function of the control voltage of the laser intensity, for the three values of $d_\mathrm{sq}$.
The dashed line is a weighted linear fit to these points, which constitutes the calibration of the potential height.


\begin{figure}[t]
\includegraphics{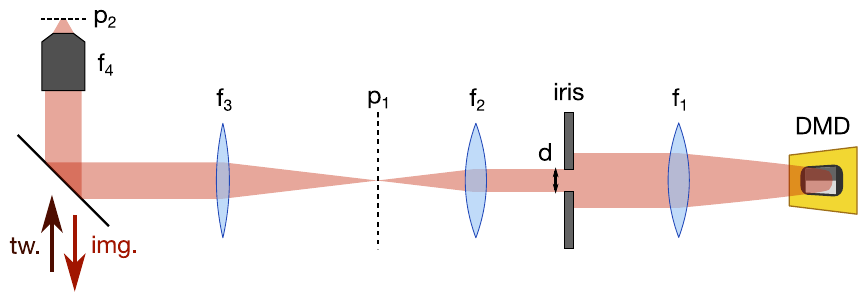}
\vspace{-0.cm}
\caption{\textbf{Optical setup to vary the edge width.}
The DMD is imaged on the atomic plane ($\mathrm{p}_2$).
An iris of variable diameter $d$ is added in the Fourier plane to vary the width of the potential step.
The DMD light at $638\,$nm is overlapped with the tweezer beam (tw.) at $1064\,$nm with a dichroic plate and focused with a microscope objective. The imaging beam (img.) at $767\,\si{\nano\meter}$ is propagating in the opposite direction and passing through the dichroic plate to reach the camera (not shown).
}
\label{Fig_SM6}
\end{figure}

\subsection{DMD light source and optical setup}
\label{sec:DMD_setup}

The DMD (ViALUX  V-7000) is illuminated with an incoherent light source 
centered at $638\,\si{\nano\meter}$ with a $3\,\si{\deci\bel}$ bandwidth of $1\,\si{\nano\meter}$ 
derived from four laserdiodes (Ushio HL63623HD) that are combined with a beam combiner module from Lasertack.
The synthesized beam is intensity stabilized using an acousto-optic modulator (Crystal Technology 3200-125) 
correcting for intensity fluctuations behind a square-core fiber (Thorlabs FP150QMT).
The square-core fiber maps the temporal incoherence into rapidly varying spatial incoherence 
thus reducing the speckle contrast to $\approx3\,\si{\percent}$.

For the measurements shown in \figref{Fig_1}-\ref{Fig_3}, 
the illuminated DMD is imaged onto the atomic plane with a single telescope 
that consists of one lens with focal length $1000\,\si{\milli\meter}$, 
and the microscope objective with focal length $f_4 = 25\,\si{\milli\meter}$.
This setup images a single pixel of the DMD (edge length $13.7\,\si{\micro\meter}$) 
onto a hypothetical patch with length $0.33\,\si{\micro\meter}$ in the plane of the atoms. 
The theoretical resolution limit of the imaging system ($\text{NA}=0.5$) 
is $\lambda/(2 \text{NA}) = 638\,\si{\nano\meter}$.
We thus average several DMD pixels for every point spread function of the objective.
With this setup, the Fourier plane of the DMD pattern is close to the microscope objective.
Since this objective is also used to focus the tweezer beam and to image the atoms 
onto the camera, reducing the sharpness with this setup is not feasible.

For the measurements presented in \figref{Fig_4} we therefore use a 
modified two-telescope setup, as illustrated in \figref{Fig_SM6}.
The DMD pattern is now imaged onto the atomic plane ($\mathrm{p}_2$) via an additional telescope.
The first one has lenses with focal lengths $f_1 = 150\,\si{\milli\meter}$ and $f_2 = 100\,\si{\milli\meter}$.
To vary the sharpness of the projected pattern, an iris with diameter $d$ is introduced in the Fourier plane after the DMD behind lens f$_1$.
The resulting image can be monitored in the image plane ($\mathrm{p}_1$ in \figref{Fig_SM6}) with a camera.
This intermediate image is then projected onto the atoms with a lens of focal length $f_3 = 750\,\si{\milli\meter}$ and a microscope objective of focal length $f_4 = 25\,\si{\milli\meter}$.
For large diameters $d$ of the iris, the finite aperture of the objective cuts the highest Fourier components and limits the sharpness of the displayed pattern.

\begin{figure}[h!]
\includegraphics{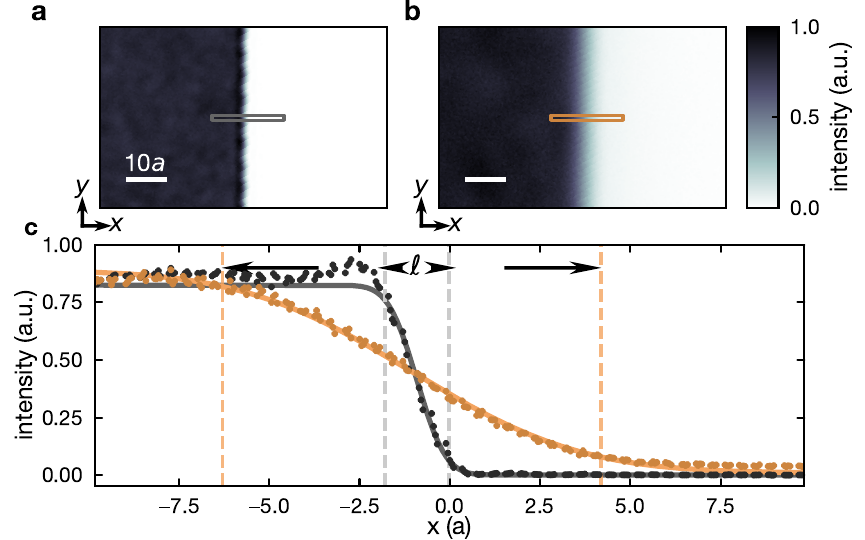}
\vspace{-0.cm}
\caption{\textbf{Optical potential created by the DMD and edge width evaluation.}
\textbf{a}, Optical potential in the intermediate plane ($\mathrm{p}_1$ in Fig.~\ref{Fig_SM6}) generated with the DMD and recorded with a camera.
The diameter of the iris, and thus the Fourier plane, is slightly larger than the corresponding size of the objective. The gray rectangle indicates the area integrated for the curve shown in \textbf{c}. 
\textbf{b}, Same as \textbf{a}, but the diameter of the iris is closed as much as possible. The brown rectangle indicates the area integrated for the curve shown in \textbf{c}.
The scale bar in \textbf{a} and \textbf{b} is $10a$ in the atomic plane $\mathrm{p}_2$.
\textbf{c}, Resulting edge width.
The edge width is extracted for a position close to the atomic cloud. We show the data of the integrated signal highlighted by the rectangles in \textbf{a} and \textbf{b} together with the corresponding fit. 
The iris diameter corresponds to a diameter slightly larger than the diffraction limit (\textbf{a}) and the smallest iris diameter from \figref{Fig_4}b (\textbf{b}).
}
\label{Fig_SM11}
\end{figure}

Two of the resulting potentials are shown as examples in \figref{Fig_SM11}a and b. We image the potential in the intermediate plane with a CMOS-camera (exposure duration $44.8\,\si{\micro \second}$) and scale the image with the demagnification of this plane.
The extracted edge width $\ell$ is illustrated in \figref{Fig_SM11}c, showing the image integrated along the edge which is highlighted by the rectangles in \figref{Fig_SM11}a and b~\cite{DefinedLength92}.
The iris diameter is set such that it is slightly larger than the size corresponding to the objective aperture in \figref{Fig_SM11}a and closed as much as possible in \figref{Fig_SM11}b.
We extract the width by fitting \eqref{eq:edge_potential} to the data, extracting $\ell = 1.8a = 0.51\,\si{\micro\meter}$ (\figref{Fig_SM11}a)  and $\ell = 10.5a = 3.0\,\si{\micro\meter}$ (\figref{Fig_SM11}b) after multiplying with the demagnification from the intermediary plane to the plane of the atomic cloud.
The observed resolution in \figref{Fig_SM11}a is slightly smaller than the diffraction limit after the objectiv, as the iris diameter was slightly larger than the corresponding size of the objective.

\cleardoublepage

\section*{Supplementary Information}

\renewcommand{\thefigure}{S\arabic{figure}}
\renewcommand{\theHfigure}{S\arabic{figure}}
 \setcounter{figure}{0}
\renewcommand{\theequation}{S.\arabic{equation}}
 \setcounter{equation}{0}
 \renewcommand{\thesection}{S\arabic{section}}
\setcounter{section}{0}


\section*{Tweezer optical setup}
The optical tweezer is produced by deflecting a laser beam at wavelength $\lambda_\mathrm{T} = 1064\,\si{\nano\meter}$ with an acousto-optical deflector (AA Optoelectronic DTSXY-400-1064). This beam is then expanded with a telescope with lenses of focal lenses $150\,$mm  and $400\,\si{\milli\meter}$ to an approximate waist of  $7.5\,\si{\milli\meter}$, and focussed on the atomic plane with our microscope objective. There, its waist is around $1.1\,\si{\micro\meter}$, and varying the deflection angle of the beam changes the position, where the tweezer is projected onto the atoms.

\section*{Trap frequency measurements}

The frequency of the tweezer in the $xy$-plane is measured by loading atoms into the tweezer from a BEC, applying the same velocity kick as described in the Methods, and letting the atoms oscillate during a variable amount of time $t_\mathrm{osc}$ in the tweezer potential.
The tweezer is then switched off abruptly and the cloud expands in the optical dipole trap during $t_\mathrm{exp} = 0.5\,\si{\milli\second}$ for large tweezer depths and $t_\mathrm{exp} =1\,\si{\milli\second}$ for lower depths.
The center-of-mass position of the atomic cloud is measured by taking absorption images.
In order to increase the signal-to-noise ratio, we average over few experimental realizations and the center-of-mass position is evaluated by fitting a Gaussian to the averaged image.
Since the typical oscillation frequency of the tweezer is on the order of $1$-$2\,\si{\kilo\hertz}$, which is comparable to $1/t_\mathrm{exp}$, the mapping from momentum to position is justified and the mean position of the cloud $\langle\boldsymbol{r}\rangle$ is related to its center of mass velocity $\langle \boldsymbol{v}_0 \rangle$ at the end of the oscillation in the tweezer via 
\begin{equation}
\langle\boldsymbol{r}\rangle \approx \langle\boldsymbol{v}_0\rangle t_\mathrm{exp} \sin\left(\omega_\perp t_\mathrm{osc}\right).
\end{equation}

We fit the observed oscillation with a sinusoidal function, whose frequency determines the in-plane trap frequency $\omega_\perp$ of the tweezer. 
For most of the experiments reported in the main text, the frequency is set to $\omega_\perp/(2\pi) = 2.0(1)\,\si{\kilo\hertz}$, where the uncertainty represents the standard deviation of the different measurements taken over a few months. 
The only difference are the diamond data points in \figref{Fig_2}c, which have been measured with an in-plane frequency of $\omega_\perp/(2\pi)=1.3(1)\,\si{\kilo\hertz}$.

The horizontal trap frequency of the dipole trap can in principle be determined 
by observing center-of-mass oscillations as described above for the tweezer trap.
The trap frequency in this case is, however, modified by the presence of the lattice beams, 
which add an anti-confining potential, thus weakening the overall trap frequency.
Therefore, we adopt a slightly different scheme.
We load a large BEC into the static honeycomb lattice at the $\Gamma$-point of the lowest band, 
let it evolve during a variable duration, and subsequently perform a bandmapping measurement 
that allows to observe the momentum distribution of the cloud.
We find that the distribution remains centered near the $\Gamma$-point at all times, 
and a residual oscillation in the harmonic trap is observed.
This oscillation has a frequency of $\omega_\mathrm{ODT}/(2\pi) = 17(1)\,\si{\hertz}$.

The vertical frequency of the dipole trap combined with the optical lattice is measured 
by observing the vertical oscillation after releasing the cloud from a tweezer trap 
that is displaced along the $z$ direction compared to the dipole trap and lattice. 
The oscillations in the trap are measured after $7\,\si{\milli\second}$ time of flight, 
mapping velocity to position. The position of the center of mass as a function of time is fitted 
with a sinusoidal curve and we extract a vertical frequency of $2\pi \times 330(30)\,\si{\hertz}$. 


\section*{Changing the direction of the kick}

\begin{figure}[!htb]
\includegraphics[width = 0.5\textwidth]{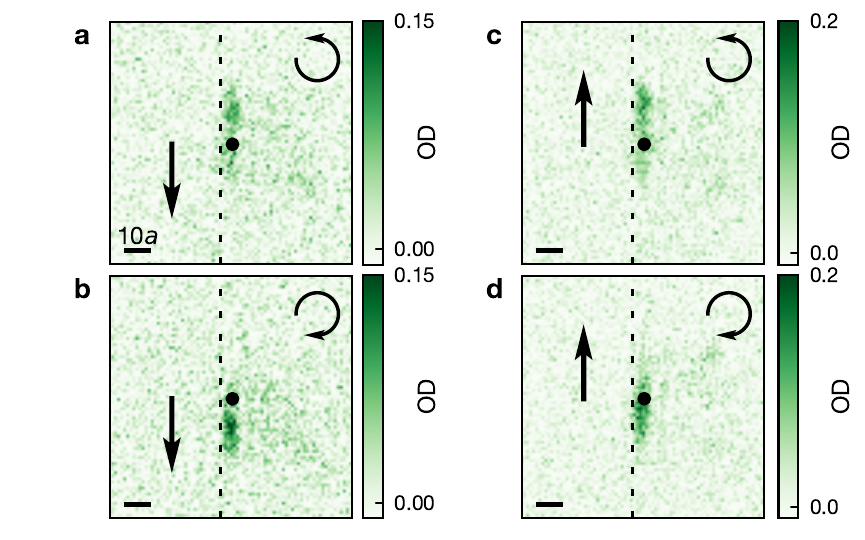}
\vspace{-0.cm}
\caption{\textbf{Changing the direction of the kick in the Haldane regime.} 
Each image \textbf{a-d} shows the evolution of the atoms with different parameters: the direction of the kick is indicated by the arrow on the left pointing up or down;
the chirality of the modulation by the direction of the circular arrow on the top right, 
corresponding to $\kappa=-1$ in \textbf{a,c}, and $\kappa=1$ in \textbf{b,d}.
The black point indicates the initial position of the cloud. 
The dashed line indicates, where the edge is located, and the scale on the bottom left represents a length of $10a$. }
\label{Fig_SM5}
\end{figure}

When preparing atoms in the edge state of the Haldane phase, a velocity kick is given to the cloud.
We verify here that the direction of this velocity kick does not change the subsequent dynamics of the cloud, in particular not its direction of propagation along the edge, which is solely determined by the chirality of the topological edge mode.
\figref{Fig_SM5} shows averaged pictures of the evolution of a cloud of atoms in the Haldane regime under the two opposite chiralities of the lattice modulation (respectively top and bottom row), and with an initial velocity kick that is applied in two opposite directions (respectively left and right column).
These pictures highlight the fact that the direction of propagation of the atoms along the edge is only determined by the chirality of the modulation, i.e., by the chirality of the topological edge mode.
The direction of the initial kick does not intervene, since the purpose of this kick is to bring the cloud of atoms from a quasi-momentum of $0$ closer to the maximum quasi-momentum of $+\pi/L_y$ with a kick in one direction or of $-\pi/L_y$ with a kick in the opposite direction.
Here $L_y$ denotes the length of the unit cell and thus $\left| \pi/L_y \right|$ is maximum possibile quasi momentum the wave packet can exhibit as these two quasi-momenta represent the same quantum state. 
The trap frequency of the tweezer is $\omega_\perp = 2\pi \times 2.0(1)\,\si{\kilo\hertz}$.


\section*{Atom number calibration}

\begin{figure}[!htb]
\includegraphics[width = 0.5\textwidth]{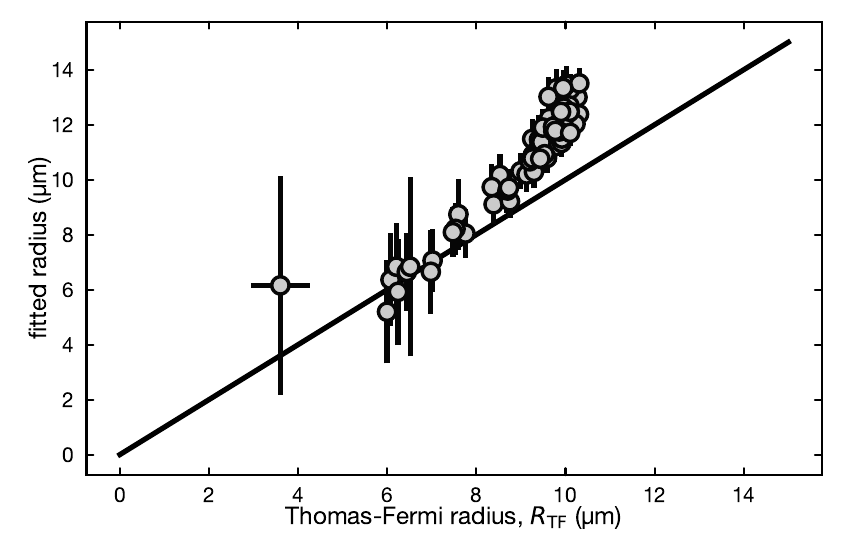}
\vspace{-0.cm}
\caption{\textbf{Atom number calibration.}
Each data point corresponds to an experimental realization of a BEC in the dipole trap.
The two estimated radial sizes of the cloud are plotted on the horizontal and vertical axes.
The horizontal axis is rescaled in order to match the estimated size of the cloud in the dilute regime.
The solid line has a slope of one, the vertical error bars represent the error of the fit of the radius, 
and the horizontal error is estimated from the background noise of the absorption image.}
\label{Fig_SM9}
\end{figure}

We calibrate the observed number of atoms in the edge mode by estimating the Thomas-Fermi radius of a small BEC in the optical dipole trap in two ways. The first one is to fit the density profile of the cloud with an inverted parabola, and the second one is to calculate it via the Thomas-Fermi formula: 
\begin{equation}
R_\mathrm{TF} = \left(\frac{15 N_\mathrm{at}a_\mathrm{s}\hbar^2\omega_z}{m_\mathrm{K}^2\omega_r^3}\right)^{1/5},
\end{equation}
where $a_\mathrm{s}$ is the $s$-wave scattering length of the atoms, $m_\mathrm{K}$ is the mass of a potassium atom, and $\omega_z$ (resp. $\omega_r$) is the vertical (resp. radial) frequency of the optical dipole trap.
In this formula, the number of atoms $N_\mathrm{at}$ is replaced by $\sigma_\mathrm{sc}\mathrm{OD}$, where $\sigma_\mathrm{sc}$ is the scattering cross-section of the imaging process, and $\mathrm{OD}$ is the summed optical density of the cloud.
The exact value of $\sigma_\mathrm{sc}$ is unknown due to the proximity of the $^2P_{3/2}$ states during the imaging, and the comparison between the two values obtained for the Thomas-Fermi radius allows to calibrate this quantity, and therefore provide the proportionality factor between the optical density and the number of atoms.

We take a series of pictures of a small BEC in the optical dipole trap, and for each picture evaluate the Thomas-Fermi radius with the two methods above.
\figref{Fig_SM9} shows the results, where the horizontal axis is rescaled by adjusting the value of $\sigma_\mathrm{sc}$ such that the Thomas-Fermi radii of the least dense clouds obtained with the two methods match: the solid line has a slope of one.
We obtain a scattering cross-section of $\sigma_\mathrm{sc} = 0.085\times 3\lambda_0^2/(2\pi)$, where $\lambda_0 = 767\,\si{\nano\meter}$ is the wavelength of the imaging light. This value is then used to estimate the number of atoms that are loaded in the edge modes.


\section*{Lifetime of the atoms in the edge state}

\begin{figure}[!htb]
\includegraphics[width = 0.5\textwidth]{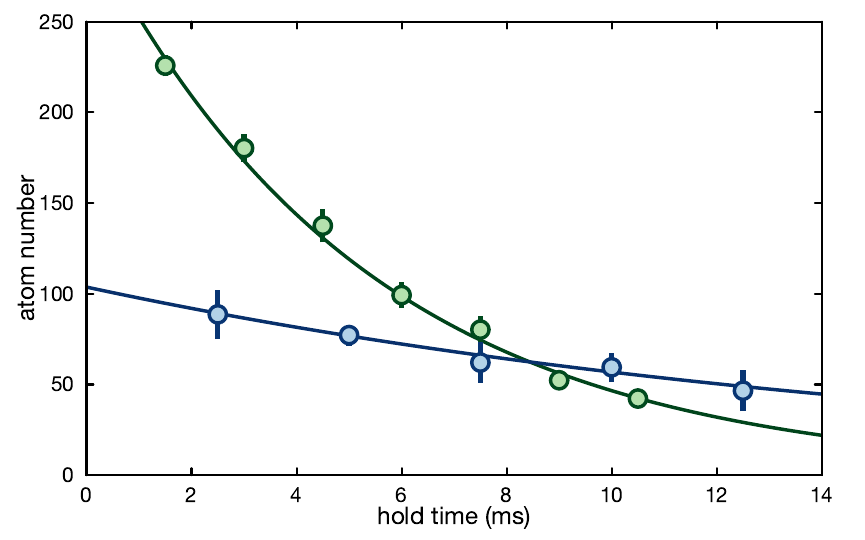}
\vspace{-0.cm}
\caption{\textbf{Lifetime of atoms in the edge state.}
The number of atoms in the edge mode is measured in two regimes: Haldane (green) and anomalous (blue).
The error bars correspond to the standard deviation of the atom numbers obtained by the bootstrapping method (see text).
The points are fitted with an exponential decay (solid lines), from which a lifetime is extracted as the only free parameter.
The presented data is an average over 10-38 averages  for the blue and 27-72 averages for the green data points, the number of averages is increasing for longer hold times.
}
\label{Fig_SM8}
\end{figure}

We measure the lifetime of the atoms in the edge state by summing the optical density of averaged pictures for various evolution times.
We evaluate the error on this sum by averaging different subsets of images and computing the summed optical density on these averages, in a similar manner to the procedure described by a bootstrapping method (see Methods section \textit{E}.
The summed optical densities and their respective error bars are then multiplied by the factor determined above to obtain the atom number.
Figure~\ref{Fig_SM8} shows the atom number as a function of hold time for two experiments, one in the Haldane regime (green), and one in the anomalous regime (blue).
We then fit each dataset with an exponential decay to evaluate the characteristic lifetime.
In the Haldane regime this lifetime is $5\,\si{\milli\second}$, and in the anomalous phase $17\,\si{\milli\second}$.
For Reference, the lifetime of a bulk BEC in the modulated lattice without any edge potential is measured to be around $100\,\si{\milli\second}$.
The reduced lifetime in the edge mode could be due to spurious dynamics in the vertical direction:
the Rayleigh length associated with the resolution of the potential edge is around $3\,\si{\micro\meter}$, 
which is smaller than the vertical extension of the atoms prepared in the tweezer.
As a result, only part of the atomic cloud may be prepared in the edge mode and the observed loss rate would then be a combination of heating in the modulated lattice and losses along the vertical direction.
In contrast, the lifetime of the atomic cloud in the modulated lattice is measured by preparing a large BEC in the combined potential formed by the dipole trap and the honeycomb lattice, which does not suffer from any mismatch of potentials in the vertical direction.
The modulation frequency in the anomalous regime is $\omega = 2\pi \times 7\,\si{\kilo\hertz}$, in the Haldane regime $\omega = 2\pi \times 16\,\si{\kilo\hertz}$, for both regimes the modulation amplitude $m = 0.25$ and the tweezer trap frequency $\omega_\perp = 2\pi \times 2.0(1)\,\si{\kilo\hertz}$.


\begin{thebibliography}{64}
\bibitem [{Def()}]{DefinedLength92}%
  \BibitemOpen
  \href@noop {} {\bibinfo {title} {Defined as the length from 8\% to 92\% of
  the height of the edge. this width is estimated assuming a good performance
  of the imaging objective, which has a {NA}$=0.5$, however, the actual width
  in the atomic plane is most likely increased due to finite
  aberrations.}}\BibitemShut {Stop}%
\end{thebibliography}
\end{document}